\documentclass[fleqn,usenatbib]{mnras}
\usepackage{graphicx}	
\usepackage{amsmath}	
\usepackage{amssymb}	
\usepackage{CJK}
\usepackage{soul}
\usepackage{ulem}

\def\msun{{\rm M_{\odot}}}

\def\be{\begin{equation}}
\def\ee{\end{equation}}

\def\del#1{{}}

\newcommand\mj{{\,{\rm M}_{\rm J}}}
\newcommand\St{\,{\rm St}}

\usepackage[dvipsnames]{xcolor}
\newcommand{\cb}[1]{\textcolor{Magenta}{#1}}

\usepackage{soul}

\title[Discs mass by planet migration and dust drift]{Using planet migration and dust drift to weigh protoplanetary discs}

\author[Wu, Baruteau \& Nayakshin., 2023]{Yinhao Wu (吴寅昊)$^1$\thanks{yw505@leicester.ac.uk},  Cl\'{e}ment Baruteau$^2$,  Sergei Nayakshin$^{1}$\\
$^{1}$School of Physics and Astronomy, University of
  Leicester, Leicester LE1 7RH, UK\\
  $^{2}$IRAP, Universit{\'e} de Toulouse, CNRS, UPS, CNES, Toulouse, France
}

\date{Accepted XXX. Received YYY; in original form ZZZ}

\pubyear{2023}

\begin{document}
\begin{CJK*}{UTF8}{gbsn}  
\label{firstpage}
\pagerange{\pageref{firstpage}--\pageref{lastpage}}
\maketitle

\begin{abstract}   
ALMA has spatially resolved over 200 annular structures in protoplanetary discs, many of which are suggestive of the presence of planets. Constraining the mass of these putative planets is quite degenerate for it depends on the disc physical properties, and for simplicity a steady-state is often assumed whereby the planet position is kept fixed and there is a constant source of dust at the outer edge of the disc. Here we argue against this approach by demonstrating how the planet and dust dynamics can lift degeneracies of such steady-state models. We take main disc parameters from the well-known protoplanetary disc HD 163296 with a suspected planet at $R\approx 86$~au as an example. By running gas and dust hydrodynamical simulations post-processed  with dust radiative transfer calculations, we first find steady-state disc and planet parameters that reproduce ALMA continuum observations fairly well. For the same disc mass, but now allowing the planet to migrate in the simulation, we find that the planet undergoes runaway migration and reaches the inner disc in $\sim 0.2$ Myr. Further, decreasing the disc mass slows down planet migration, but it then also increases the dust's radial drift, thereby depleting the disc dust faster. We find that the opposing constraints of planet migration and dust drift require the disc mass to be at most $0.025~\msun$, must less massive than previously estimated, and for the dust to be porous rather than compact. We propose that similar analysis should be extended to other sources with suspected planetary companions.
\end{abstract}


\section{Introduction}
Protoplanetary discs are the engines that drive formation of young protostars as well as formation and early orbital evolution of planetary systems. Mass is one the most fundamental characteristics
of protoplanetary discs, yet it remains very difficult to constrain observationally. This is mainly because H$_2$, by far the most dominant constituent of protoplanetary discs, emits very weakly in the range of temperatures encountered at radii where most of the discs mass is contained \citep{Carmona10}. Consequently, one needs to rely on two main approaches to estimate the disc mass: emission by minor constituents (gas species or dust), and discs dynamics.

In the first approach, dust particle continuum emission at mm-wavelengths has been used widely to estimate disc mass. Provided that dust emission is optically thin, that dust scattering is negligible, that the dust's absorption opacity and temperature are uniform and known, disc dust mass can be inferred from the averaged flux of dust continuum emission for a source with known distance to Earth. The total mass of the disc can then be obtained by further assuming the dust-to-gas mass ratio (see \citealp{Miotello22-PPVII} for a recent review). Detailed modelling shows that uncertainties in the assumptions mentioned above supplemented by modelling freedom in dust particle size distribution, composition and porosity \citep[e.g.,][]{Kataoka14-porous-dust, Woitke16-DIANA}, result in order of magnitude uncertainties in the inferred disc masses \citep[e.g., see][]{Andrews18-Rdisc-dust, Zhu19-scattering-albedo,Miotello22-PPVII,Xin22-dust-masses-opacities}. 

The emission of minor gas constituents, like CO and its isotopologues, is an alternative route to tracing the discs mass, but it also has its own flaws, including optical depth effects and freeze-out onto dust grains \citep{BerginEtal13, Kamp17-CO}. Because it does not freeze out and its abundance relative to H$_2$ is well known, HD is a promising disc mass tracer despite the degeneracy between its abundance and the gas temperature \citep{BerginEtal16, TrapmanEtal17, Schwarz22}, however only a handful protoplanetary discs have HD emission data currently. All in all, the estimation of disc masses based on dust and/or gas emission is probably uncertain by an order of magnitude \citep{Miotello22-PPVII}.

In the second approach based on discs dynamics, several methods have been proposed. First, the disc mass can be crudely estimated via the observed stellar accretion rate ($\dot M_{\star}$) and the age of the star ($t_{\star}$) as $M_{\rm disc} \approx \zeta \dot M_{\star} t_{\star}$, where $\zeta$ is a few \citep{JPA12}. This assumes a quasi-steady state evolution for the disc gas driven by radial (turbulent) transport of angular momentum, an assumption that is more and more challenged by the emergent picture of vertical transport due to magnetised winds \citep{Suzuki16-MHD-winds, Lesur21-MHD-winds, Tabone22-general, MHD-wind-Elbakyan, wu2023wind}. A second dynamics method is based on dynamics of dust particles rather than that of gas. So long as dust particles have a Stokes number $\St \ll 1$, their radial drift velocity due to gas drag is directly proportional to the dust particle size ($s$) and inversely proportional to the gas density \citep{BirnstielEtal12}. By observing a disc at different wavelengths, one could in principle infer the radial location of the outer edge of the dust in the disc as a function of $s$, which can then be turned into an estimate of the gas disc mass \citep{Powell17-dust-lines, Powell19-DustLanes}. Unfortunately, uncertainties in the dust opacity, and the fact that dust porosity strongly affects the radial dust drift rates at high porosity, preclude this method from being sufficiently constraining \citep{franceschi22-dust-lanes}. Note also that both aforementioned methods assume the age of the system is known, while it is usually poorly constrained for young stellar objects. Another method measures departure of a disc rotation curve from its Keplerian profile, but this is well suited only to very massive self-gravitating discs \citep{Veronesi21}.

In this paper, we propose another dynamical method to estimate the discs mass. It is based on planetary migration in discs for which dust sub-structures, in particular dark rings in the dust continuum emission, are assumed to have a planetary origin. In less than a decade, spatially resolved observations of protoplanetary discs have uncovered a wealth of sub-structures in both dust and gas emission that much resemble those seen in simulations of disc-planet interactions \citep[e.g.,][]{BroganEtal15, Dsharp2, Dsharp7, LongEtal18, Casassus19-HD100546, TsukagoshiEtal19, Perez2019, Pinte20-Dsharp-Vkinks}. The planetary interpretation of these structures can (and should) be debated as other planet-free mechanisms may also form similar structures in the discs emission \citep{Meru17,Riols20}. Still, if discs structures are due to planets, then it is possible to constrain the mass of these planets by comparing synthetic images of the discs emission with real observations. In doing this comparison, synthetic images are most often based on gas and dust hydrodynamical simulations of disc-planet(s) interactions in which planetary migration is discarded for simplicity \citep[e.g.,][]{Dsharp7}. This assumption is valid if planet migration timescale is long in comparison to the characteristic timescale for  evolution of gas and dust in the disc, a lower limit of which being set by dust particles radial drift. Since the speed of planetary migration increases with both planet mass and disc mass (or, actually, with the disc's surface density), assuming that disc structures are caused by planets on fixed orbits may be reasonable for some system parameters but not for others. A recent illustration of this point has been presented in \cite{TypeIII-Wafflard-Fernandez2020}. They found that a single planet, with a mass typically between that of Saturn and Jupiter, can undergo rapid intermittent type III migration provided that the disc is massive enough. This unsteady migration pattern may trigger formation of multiple dust rings and gaps which in a fixed planet orbit scenario would require multiple planets.
It is also known that migrating planets produce different effects on the structure of the gaps and rings they carve in the disc compared with static planets \citep{Meru19-Ring,Perez2019,Weber2019-dust-diffusion}. It is possible that dust continuum emission in planet vicinity can serve as a handle on the unknown disc mass.

Furthermore, in this paper we point out  the following simple constraint that planet migration and dust dynamics place on the disc mass when taken together. It is well known that even for low mass planets, in the Neptune to Saturn mass range, the migration timescale may be quite a small fraction of the disc lifetime $t_\star$ at high disc masses \citep{BaruteauEtal14a}. On the other hand, to have a decent statistical chance of catching a migrating planet in the process of structuring its dusty disc we would expect its migration timescale, $t_{\rm m}$, to be a fair fraction of the disc age. Indeed, $t_{\rm m}$ is the rough measure of how long it takes for the planet to disappear from the outer disc into the inner unresolved one, so if we happen to image a planet in the outer disc then it is fair to assume that it is not older than $\sim t_{\rm m}$. But if planets form at random times with respect to the time when we image the discs then the probability of observing a given planet is $\sim t_{\rm m}/t_\star$. We thus argue that a maximum mass for structured discs can be set in order to avoid too fast planetary migration. While only an upper disc mass can be set with this approach, it may turn out quite useful in conjunction with the other methods discussed above. For example,  the dust lines method of \cite{Powell17-dust-lines, Powell19-DustLanes}, gives a lower limit on the disc mass. By combining dust and planet dynamics the discs mass could hopefully be constrained from both below and above. 

The present study aims at testing these ideas via multi-fluid hydrodynamical simulations of disc-planet interactions post-processed by dust radiative transfer calculations. Our disc model is inspired by the well-known disc around the $M_{\star} \approx 2\,\msun$ Herbig Ae star HD 163296, a relatively old protostar with an estimated age $t_{\star} \approx (7-10)$~Myr \citep[e.g.,][]{Vioque-18-star-catalogue,Setterholm-18-HD163296}. The disc has been observed with ALMA as part of the DSHARP Large Programme \citep{Dsharp1, Dsharp2}. Dust continuum emission from the system features at least 4 dark rings including two prominent ones located around 48 and 86 au \citep{IsellaEtal18-HD163296}. Localised velocity perturbations based on CO gas emission strongly suggest the existence of several massive planets in the disc \citep{Teague18, Pinte20-Dsharp-Vkinks}, including one inside the dark ring around 86 au, which our study will focus on. Especially relevant to our study is the fact that the mass of the disc in HD 163296 has been estimated by various techniques: through dust continuum emission ($\approx 0.12~\msun$, \citealp{Powell19-DustLanes}), dust dynamics ($\approx 0.2~\msun$, \citealp{Powell19-DustLanes}) and also via chemodynamical modelling of the emission of various CO isotopologues and of HD (between 0.05 and 0.3$~\msun$, \citealp{BoothEtal19-HD63296, Kama20}).

This paper is organised as follows. Section~\ref{sec:methods} details our disc model inspired from the HD 163296 disc, and describes the setup of our hydrodynamical simulations and dust radiative transfer calculations. Section~\ref{sec:results} presents our results of simulations and dust radiative transfer calculations. A few concluding remarks follow in Section~\ref{sec:conclusion}.

\section{Disc model and numerical methods}\label{sec:methods}

As already stated in the introduction, here we focus on HD 163296 system as a case study. We focus specifically on the dark ring in the dust continuum emission at around 86 au that has been suggested to be carved by a massive planet \citep[e.g.,][]{Liu18-HD163296, Juan2023-HD163296}. Detection of a localised velocity perturbation from the dark ring's vicinity \citep{Teague18, Pinte20-Dsharp-Vkinks} makes the case for a massive planet especially compelling. We emphasize that we do not aim here to reproduce all the structural features observed in the disc of HD 163296, but just use the disc parameters as an example. For a recent example of a study setting that aim we point the reader to \cite{Juan2023-HD163296}, which came out while this manuscript was in review. The authors use four massive planets held on fixed orbits in a massive gas disc to try and reproduce HD 163296 observations in detail. Again, our goal here is to to interpret the dark ring in HD 163296 as a dust gap carved by a single massive planet, and to explore what planet and disc masses (among other physical parameters) could account for the ring properties when planet migration is taken into account. To this aim, we have carried out two-dimensional (2D) hydrodynamical simulations of disc-planet interactions to model the gas and dust evolution of a disc whose physical properties are chosen to reflect those in the HD 163296 disc (see Section~\ref{sec:hd}). The dust surface density obtained in the hydrodynamical simulations is then used to calculate synthetic maps of the dust continuum emission in the sub-millimetre via 3D dust radiative transfer calculations (see Section~\ref{sec:RT}). 

Our work is split into two stages. In the first one, we assume that the planet remains on a fixed circular orbit and we examine what disc and planet parameters can account for the flux properties of the dark ring of emission near 86 au in the HD 163296 disc. Having a fixed planet is a common simplifying assumption that has often been used in previous studies to infer what planet mass would be needed to explain the width and depth of dark rings of emission in the sub-mm. In the second stage of our study, we consider the radial migration of the planet, given the disc and planet parameters obtained in the first stage, and we examine how planet migration affects our findings.

\subsection{Setup of hydrodynamical simulations}\label{sec:hd}
We use the public grid-based code FARGO3D \citep{BL_Masset_16_FARGO3D} in its multi-fluid version \citep{BL19-FARGO} to model the evolution of a protoplanetary disc with an embedded planet. The disc model includes both gas and dust, which interact via linear drag forces. Planet-disc interaction due to gravity is included, but gas self-gravity is neglected. Our simulations are 2D and in the following, we describe the disc by its polar coordinates $R$ and $\varphi$. Since the star remains at the frame's origin, the indirect terms arising from the planet and the disc accelerations are fully taken into account.

\subsubsection{Gas model}\label{sec:gas}
We assume that the evolution of the disc gas is driven by radial transport of angular momentum via a turbulent viscosity. We set the turbulent viscosity to the form of the isotropic prescription by \cite{Shakura73} with a constant $\alpha$ viscosity parameter. ALMA CO and DCO$^{+}$ molecular line observations of the HD 163296 disc only set an upper limit to the disc turbulence parameter $\alpha \lesssim 3\times 10^{-3}$ \citep{Flaherty17-HD163296-turb}. We have therefore experimented with two values of the $\alpha$ parameter: $10^{-4}$ and $10^{-3}$ \citep{Liu-2022}.

Because the total mass in the HD 163296 disc is not constrained accurately neither, we have also varied the disc mass in our simulations, which allowed us to study its impact on the planet migration rate as well as on the dust drift rate. The disc mass $M_{\rm disc}$ ranges from 0.0125~$M_\odot$ to 0.2~$M_\odot$. Following \cite{Dullemond20-HD163296}, the initial gas surface density profile at time $t=0$ in the simulation is given by
\begin{equation}
    \Sigma_{\rm g}(R, t = 0) = \Sigma_{\rm g, 0} \left(\frac{R}{R_0}\right)^{-p} \exp\left[ -\left(\frac{R}{R_{\rm exp}}\right)^{{2-p}} \right]\;,
    \label{sigma-disc0}
\end{equation}
with $R_0 = 100 $~au, $R_{\rm exp} = 150$~au, $p = 1$, and $\Sigma_{\rm g, 0}$ is a normalisation constant that changes with $M_{\rm disc}$.

A locally isothermal equation of state is adopted for simplicity, meaning that the gas temperature stays stationary. This assumption is most presumably valid given that we model regions in the HD 163296 disc located around 100 au, where the cooling timescale is expected to be much shorter than the dynamical timescale. We set the radial profile of the gas temperature, $T(R)$, as
\begin{equation}
    T(R) \simeq 15 \left (\frac{R}{R_{0}} \right)^{-0.14}~K,
    \label{temperature}
\end{equation}
which features the same power-law exponent as in the midplane temperature inferred from \cite{Dullemond20-HD163296} by the modelling of CO($2-1$) emission in the disc. Our reference temperature at $R_{0} = 100$ au is, however, slightly smaller than their midplane temperature at the same location ($\sim$25 K). This is motivated by preliminary simulations and dust radiative transfer calculations, which showed that, for the range of planet masses that we have experimented, a smaller disc temperature was required to obtain a narrower gap around the planet's location, closer to that in the ALMA continuum observation (see also \citealp{dong2018eq10}). This is similar to the simulations presented in \citet{Perez2019}, where a disc temperature quite smaller than what would be usually assumed was necessary for a gap-opening planet to account for the short radial separation between the three fine bright rings in the outer parts of the disc around HD 169142 as observed with ALMA. Assuming a Solar-like composition for the disc gas (mean molecular weight equal to 2.4), our temperature profile translates into the following radial profile for the disc's aspect ratio, $h(R)$:
\begin{equation}
    h(R) = \frac{c_{\rm s}}{v_{\rm K}}(R) = 0.054\times \left(\frac{R}{R_0} \right)^{0.43},
    \label{aspect ratio}
\end{equation}
where $v_{K}$ is the Keplerian velocity in which we have assumed the stellar mass to be $M_{\star} = 2.0~\msun$ \citep{Dsharp1}.

\subsubsection{Dust model}\label{sec:dust}
Dust is modelled as multiple pressure-less fluids, and our simulations include dust diffusion \citep{Weber2019-dust-diffusion} as well as dust feedback on the gas. We have modified the public source files of FARGO3D such that each dust fluid corresponds to a fixed particle size instead of a fixed Stokes number. The simulations presented in this paper use 5 dust fluids with size $s$ spaced logarithmically uniformly between 30 $\mu$m and 3 mm. We adopt an initial size distribution in $s^{-3.5}$ and an initial dust-to-gas mass ratio (and surface density ratio) that defaults to 1\%. In the Epstein regime, which is relevant to our disc model, the Stokes number can be approximated as 
\begin{equation}
    \St \approx \pi/2 \times s\rho_{\rm d}/\Sigma_{\rm g},
    \label{stokes_number}
\end{equation} 
where $\rho_{\rm d}$ is the internal mass volume density for each dust fluid and $\Sigma_{\rm g}$ is the local gas surface density. By default, our simulations use $\rho_{\rm d}=1.3~\rm{g\,cm}^{-3}$, which is also consistent with the dust composition assumed in our dust radiative transfer calculations. Simulations with smaller internal densities will be presented in Section~\ref{sec:dust-result}.

\subsubsection{Planet setup}\label{sec:planet}
\cite{dong2018eq10} showed how the radial location of the maxima and minima in the dust's surface density around a gap-opening planet depends on the planet mass $M_{\rm p}$ and the disc's aspect ratio for inviscid discs. Following these authors, and given our disc's aspect ratio, we estimate that $M_{\rm p}$ should be $\approx 0.26 \mj$ to approximately match the radial width of the dark ring in the mm-emission that is located around 86~au in the HD 163296 disc. We therefore set the planet-to-star mass ratio to $1.3\times10^{-4}$. Given that this value is inferred from inviscid disc models, while our simulations do include viscosity, we also tried lower and higher planet masses (see Appendix \ref{planet-mass}). In order for the disc to accommodate the disturbance generated by the planet smoothly, we let the planet's mass increase from zero to its final mass according to the formula
\begin{equation}
    M_{\rm taper} = M_{\rm p} \sin^{2} \left (\frac{\pi}{2} \times\frac{t}{t_{\rm taper}} \right)\;,
    \label{mass_taper}
\end{equation}
with $t_{\rm taper} = 10$~orbits is the timescale for the mass of planet to increase from 0 to $M_{\rm p}$.

In our simulations, we either hold the planet on a fixed circular orbit at 86 au or we let it migrate freely in the disc. In the latter case, we consider two cases: (i) the planet starts migrating at $R_{\rm p, start}$=86 au after having remained on a fixed orbit for 100 orbits, and (ii) the planet starts migrating at $R_{\rm p, start}$=130 au after having remained on a fixed orbit for only 10 orbits. Since neither gas nor dust self-gravity is included in our simulations, the total disc's surface density is subtracted from its azimuthal average prior to the calculation of the disc force on the planet \citep{BM08,BL_Masset_16_FARGO3D}, and the force exerted by the gas and dust inside the planet's Hill radius is discarded \citep{Crida2009-ExcludeHill}. To mimic the effects of finite vertical thickness, the planet's gravitational potential is smoothed over a softening length set to 0.6 pressure scale heights at the planet's orbital radius \citep{smooth-length2012}.

\subsubsection{Numerical setup}\label{sec:setup}
We use 512 grid cells in the radial direction with a linearly uniform radial spacing, and 1536 in the azimuthal direction. The grid extends from 0.3 $R_0$~(30 au) to 2.2 $R_0$~(220 au) in the radial direction, and from 0 to $2\pi$ in the  azimuthal direction. Regarding boundary conditions, for the gas we use so-called wave-damping boundary conditions, whereby the gas density and velocity components are damped towards their initial radial profile. For the dust fluids, we use an open boundary condition for their radial velocity at the grid's inner edge, and we do not use wave-damping. This implies that the surface density of the dust fluids at the grid's outer edge \cb{is} not kept constant over our simulations, which means there is a finite supply of dust from outside of the system.

\subsection{Setup of dust radiative transfer calculations}\label{sec:RT}
Our dust radiative transfer calculations were carried out with the public code RADMC-3D \citep{RADMC-3D-2012}. The post-processing of our FARGO3D simulations as inputs to RADMC-3D, and the calculation of the final beam-convolved synthetic maps of dust continuum emission were done with the public python code \texttt{fargo2radmc3d}. The setup of the dust radiative calculations is very similar to that described in \cite{Baruteau21}, in particular the fact that the dust temperatures used in RADMC-3D correspond to the temperature used in the hydrodynamical simulations. We therefore only briefly list the main parameters used. The surface density of each of our 5 dust fluids is projected onto a 3D spherical grid assuming vertical hydrostatic equilibrium and dust scale heights set to 
\begin{equation}
    h_{\rm i,dust}=h(R)\times\left(\frac{\alpha}{\alpha+\St_{\rm i}}\right)^{1/2},
    \label{dust-scale-height}
\end{equation}
where $i=1..5$ and $\St_{i}$ is the average Stokes number for the $i^{\rm th}$ dust fluid. The spherical grid covers about $\pm$3H over the midplane over 40 cells with a non-uniform spacing. Henyey-Greenstein anisotropic dust scattering is included. The dust absorption and scattering opacities are calculated according to Mie theory, assuming a default dust composition comprised of 70\% water ices and 30\% astrosilicates, thereby corresponding to an internal mass volume density of about 1.3 g cm$^{-3}$ (the same value as in the hydrodynamical simulations). Note that the results of hydrodynamical simulations and dust radiative transfer calculations presented in Section~\ref{sec:dust-result} were obtained with porous dust having an internal density of 0.1 g cm$^{-3}$, assuming the same composition as above but an additional volume fraction of vacuum set to $\sim92$ per cent \citep{fargo2radmc3d-dust}.

The specific intensity of the dust continuum emission is computed at a wavelength of 1.25 mm with $10^8$ photon packages used for ray-tracing. The disc is taken to have a distance of 101 pc, an inclination of 46.7 degrees and a position angle of 133 degrees. To produce the final synthetic image of the disc, we first add white noise to the map of specific  intensity computed by RADMC-3D (prior to beam convolution) with an rms noise level of $2.3 \times 10^{-5} \rm Jy/beam$, and the obtained map is finally convolved with an elliptical beam of $0.048\times0.038$~arcsec$^2$ and position angle of 81.7 degrees \citep{IsellaEtal18-HD163296}.

\section{Results}\label{sec:results}

\subsection{Simulations Table and guide to main results}

\begin{table*}\textbf{{Hydrodynamical simulations models: parameters, results and comments}}
\centering
\begin{tabular}{|c|c|c|c|c|c|c|c|c|c}\hline
Model&$M_{\rm disc}$&$M_{\rm p}$&$\rho_{\rm d}$&Migration&$R_{\rm p, start}$&$t_{\rm m}$&Comments\\
 &$(M_{\odot})$&$(M_{\rm J})$&$(\rm{g\,cm}^{-3})$&(Yes/No)&(au)&(kyr)& 
\\\hline
\hline
NM05&0.05&0.26&1.3&No&86&&best-fit model w/o planet migration\\\hline
NM05a4&0.05&0.26&1.3&No&86&&$\alpha = 10^{-4}$, non-axisymmetric rings\\\hline
NM025&0.025&0.26&1.3&No&86&&too wide gap\\\hline
NM05H&0.05&0.52&1.3&No&86&&too wide gap (Appendix \ref{planet-mass}) \\\hline
NM05L&0.05&0.13&1.3&No&86&&good fit model (Appendix \ref{planet-mass})\\\hline
\hline
OM0125&0.0125&0.26&1.3&Yes&86&3612&slow migration\\\hline
OM025&0.025&0.26&1.3&Yes&86&1868&slow migration\\\hline
OM05&0.05&0.26&1.3&Yes&86&425&runaway migration\\\hline
OM1&0.1&0.26&1.3&Yes&86&190&runaway migration\\\hline
OM2&0.2&0.26&1.3&Yes&86&118&runaway migration\\\hline
\hline
PM025&0.025&0.26&1.3&Yes&130&1985&no outer bright ring\\\hline
PM05&0.05&0.26&1.3&Yes&130&869&weak outer bright ring\\\hline
PM05H&0.05&0.52&1.3&Yes&130&538&too wide gap (Appendix \ref{planet-mass})\\\hline
PM05L&0.05&0.13&1.3&Yes&130&1282&no outer bright ring (Appendix \ref{planet-mass})\\\hline
\hline
PM025p&0.025&0.26&0.1&Yes&130&2006&insufficient flux of continuum emission\\\hline
PM025d10p&0.025&0.26&0.1&Yes&130&2006&$\epsilon = 0.1$, best-fit model w/ planet migration\\\hline
\end{tabular}
\caption{\label{table1}List of models adopted in the hydrodynamical simulations. The main parameters are listed in Columns 2 to 6: the disc mass $M_{\rm disc}$, the planet mass $M_{\rm p}$, the internal mass volume density $\rho_{\rm d}$ for the dust fluid, whether the planet is allowed to migrate or not, and the initial orbital radius of the planet. Column 7 indicates the migration timescale $t_{\rm m}$ Eq.~(\ref{migration-timescale}) as measured from the simulations. The last column uses a comment to characterise main results of the simulations.}
\end{table*}

In this section we present results of our hydrodynamical simulations and of dust radiative transfer calculations. For reader's convenience in Table \ref{table1} we summarise the parameters and key results of our simulations. The Table is broken into four horizontal blocks separated by double lines, with model names reflecting one or more major model assumptions. In the first section we present the case of a planet on a fixed orbit at 86 au (Section~\ref{sec:simovserve-fixed}). These model names start with ``NM" to emphasise that they include no migration for the planet. The purpose of this set of models is to constrain the disc's mass and turbulent viscosity that yield synthetic images in a reasonable agreement with the observed ones. Having constrained in this way viscosity parameter, $\alpha = 10^{-3}$, we use this value for the rest of the paper.

In Section~\ref{sec:migration-86au} we then relax the assumption of a fixed orbit for the planet, allowing it to migrate freely. The names of these models in Table 1 start with ``O'' to indicate that the planet in our model starts from the ``observed" position of the dark ring of continuum emission at 86~au \citep{Dsharp2}. The models in this section differ by the disc mass, which we vary in steps of a factor of 2 from the low value of $M_{\rm disc} = 0.0125\,\msun$ in Model OM0125 to the high value of $M_{\rm disc} = 0.2\,\msun$ in Model OM2. The main purpose of these experiments is to demonstrate how planet migration can constrain the otherwise unknown disc mass $M_{\rm disc}$; in the case of HD 163296 we find $M_{\rm disc} \leq 0.025\,\msun$ for our fiducial planet mass. This value for the disc mass challenges the conclusion of the fixed planet simulations in Section~\ref{sec:simovserve-fixed}, where we find a good match to the data with $M_{\rm disc} = 0.05\,\msun$. 

For a more comprehensive comparison of our simulations with the dust continuum observations, in Section~\ref{sec:migration-130au} we let the planet to start much further out, at $R_{\rm p, start}=130$~au and run the simulations until the planet reaches $R_{\rm p, end}=86$~au. The names of these models in Table 1 start with ``P'' to indicate that the planet starts from a ``past" position -- it is unlikely that the planet currently observed at 86 au formed there without undergoing migration from further out. Similar to  Section~\ref{sec:simovserve-fixed} we perform dust radiative transfer calculations to build synthetic images when the planet reaches 86 au. 
The comparison of the synthetic images with the observations shows that there is no disc mass that simultaneously satisfies the opposing constraints of dust drift and planet migration. At high disc masses the planet reaches the inner parts of our computational grid too rapidly, whereas at low disc masses the planet migration is reasonably slow but the dust reaches the inner parts of the grid too rapidly. This conclusion therefore holds whether we start the planet at 86 au or let it migrate from further out.

Therefore, in Section~\ref{sec:dust-result} we propose that the dust in the HD 163296 disc may be more porous than assumed in Sections~\ref{sec:simovserve-fixed}--\ref{sec:migration-130au}. More specifically, we set the internal dust density to $\rho_{\rm d}=0.1$~g\,cm$^{-3}$ in both the hydrodynamical simulations and the dust radiative transfer calculations, and the names of these models in Table 1 end with ``p'' to reflect the porous dust assumption. We then find that porous dust does mitigate the opposing challenges of the too rapid dust drift and planet migration but then results in an insufficient dust continuum flux in the synthetic images. We finally explore the idea that the disc in HD 163296 may be much more dust rich than usually assumed, by increasing the initial dust-to-gas mass ratio to 0.1 in Model PM025d10p.

\subsection{Reference case: a planet on a fixed orbit}\label{sec:simovserve-fixed}
The disc's turbulent viscosity is a major unknown that affects the depth of a gap opened by a planet of a given mass in a disc with a given aspect ratio. Recall that in our disc models, $h(R)$ is given by Eq.~(\ref{aspect ratio}) and our fiducial planet-to-star mass ratio is $q=1.3\times10^{-4}$. Following previous works \citep[e.g.,][]{Liu18-HD163296, Dsharp7}, we constrain the disc's $\alpha$ turbulent viscosity by comparing the shape and width of the dark ring in dust continuum emission caused by the planet in our simulations with those observed in HD 163296 around 86 au. As pointed out in Section~\ref{sec:gas}, an upper limit to $\alpha$ from molecular line observations is $\sim3\times 10^{-3}$. Therefore, we tested two values of the $\alpha$ turbulent viscosity: $10^{-3}$ (models NM05 and NM025) and $10^{-4}$ (Model NM05a4).

\begin{figure*}
\centering
\includegraphics[width=0.49\hsize]{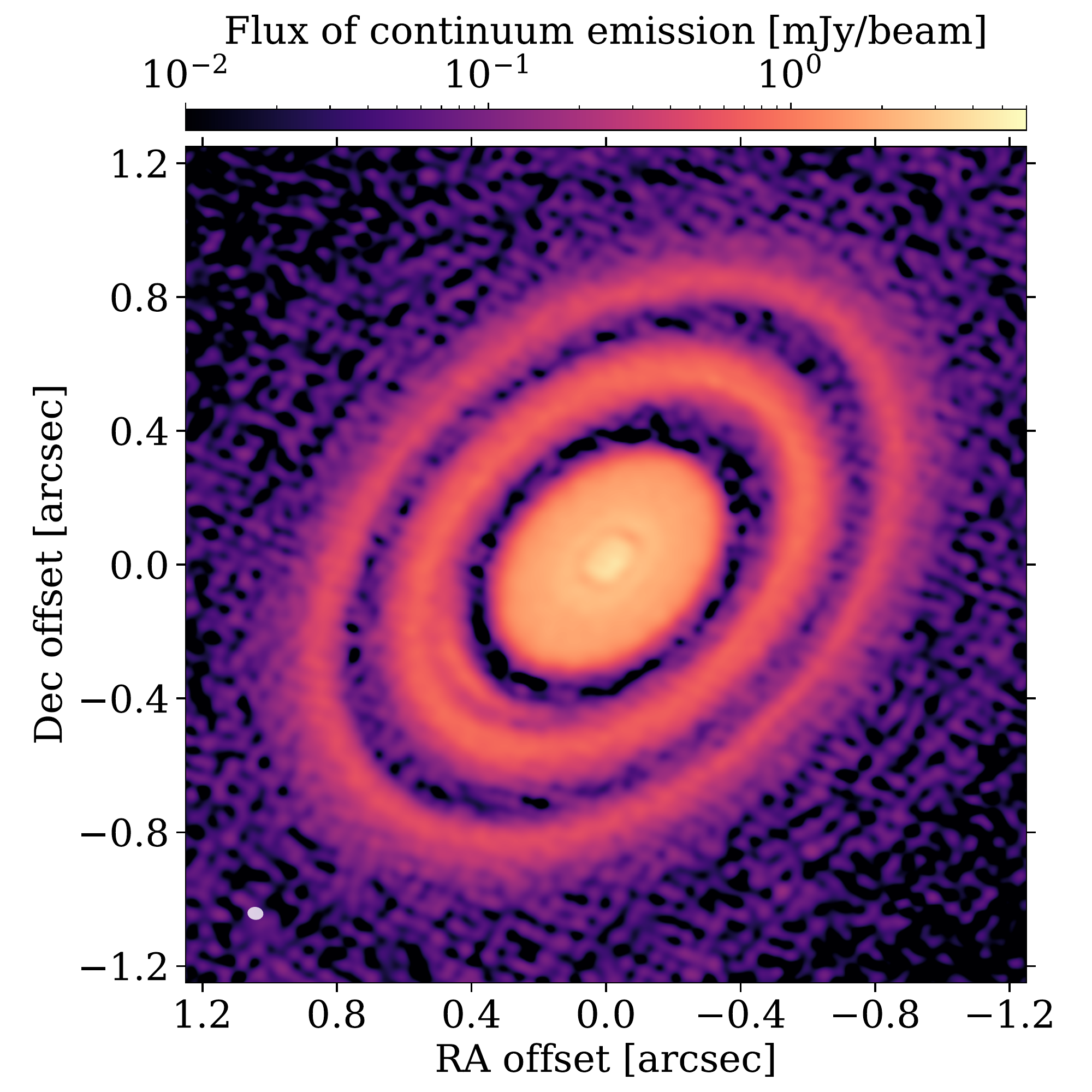}
\includegraphics[width=0.49\hsize]{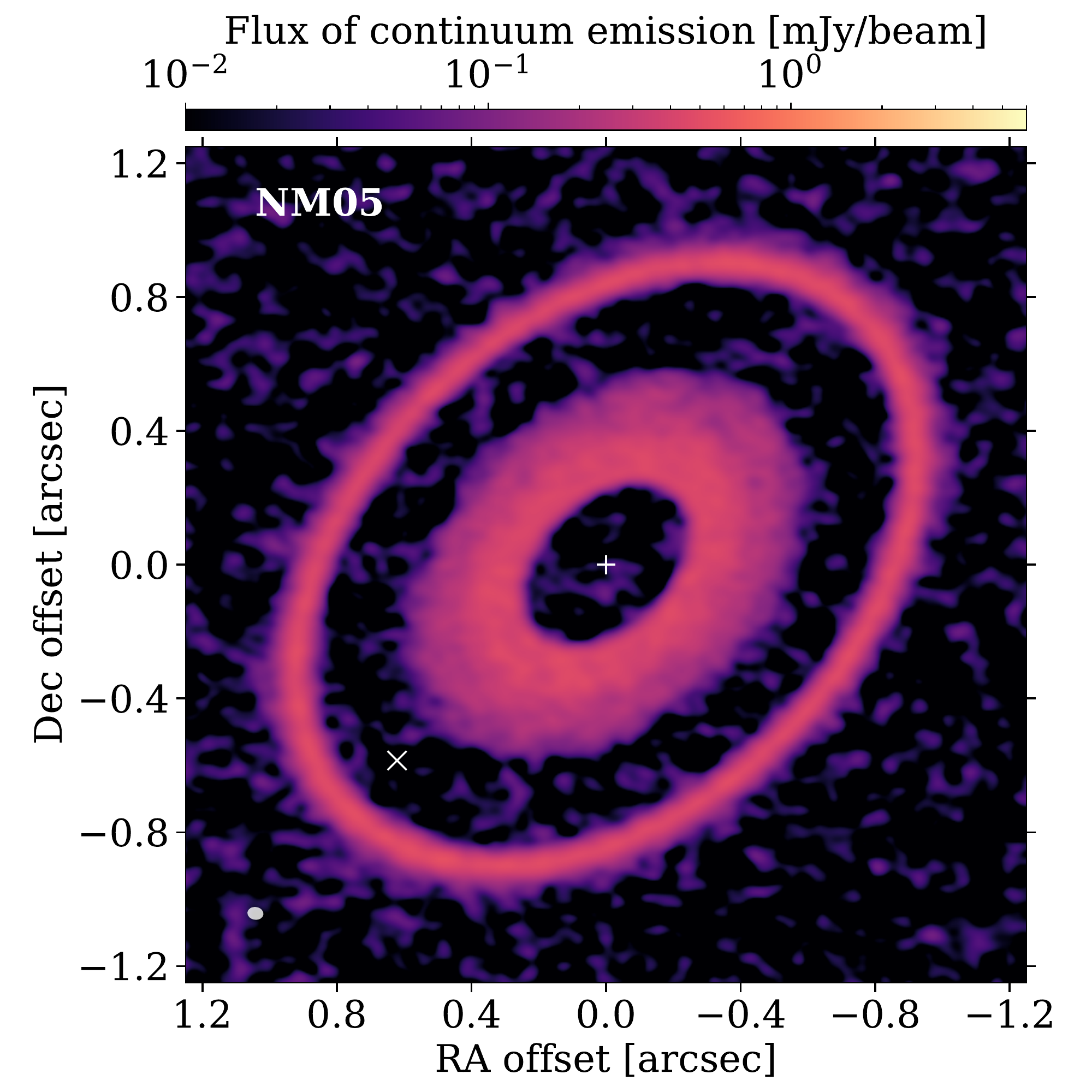}
\includegraphics[width=0.49\hsize]{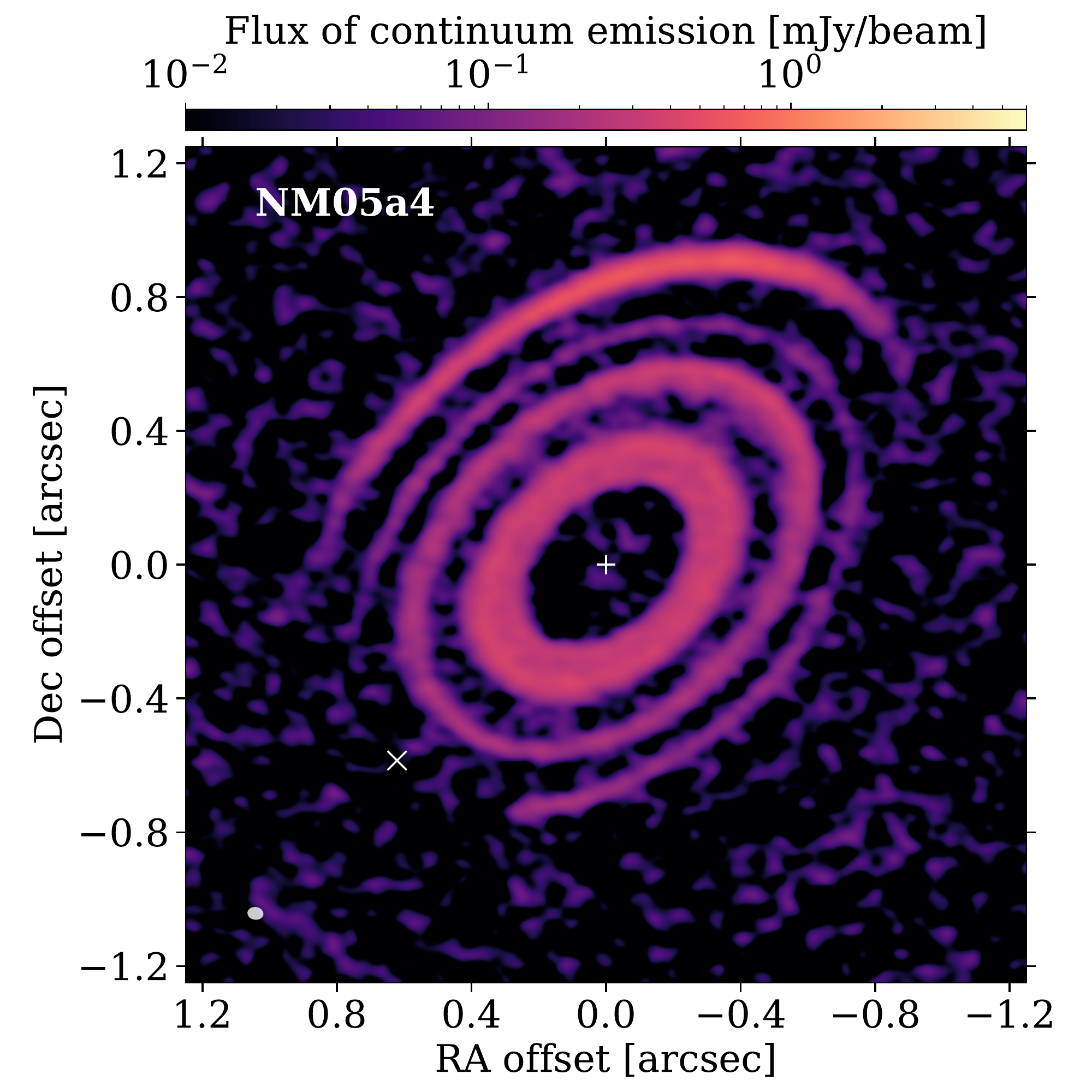}
\includegraphics[width=0.49\hsize]{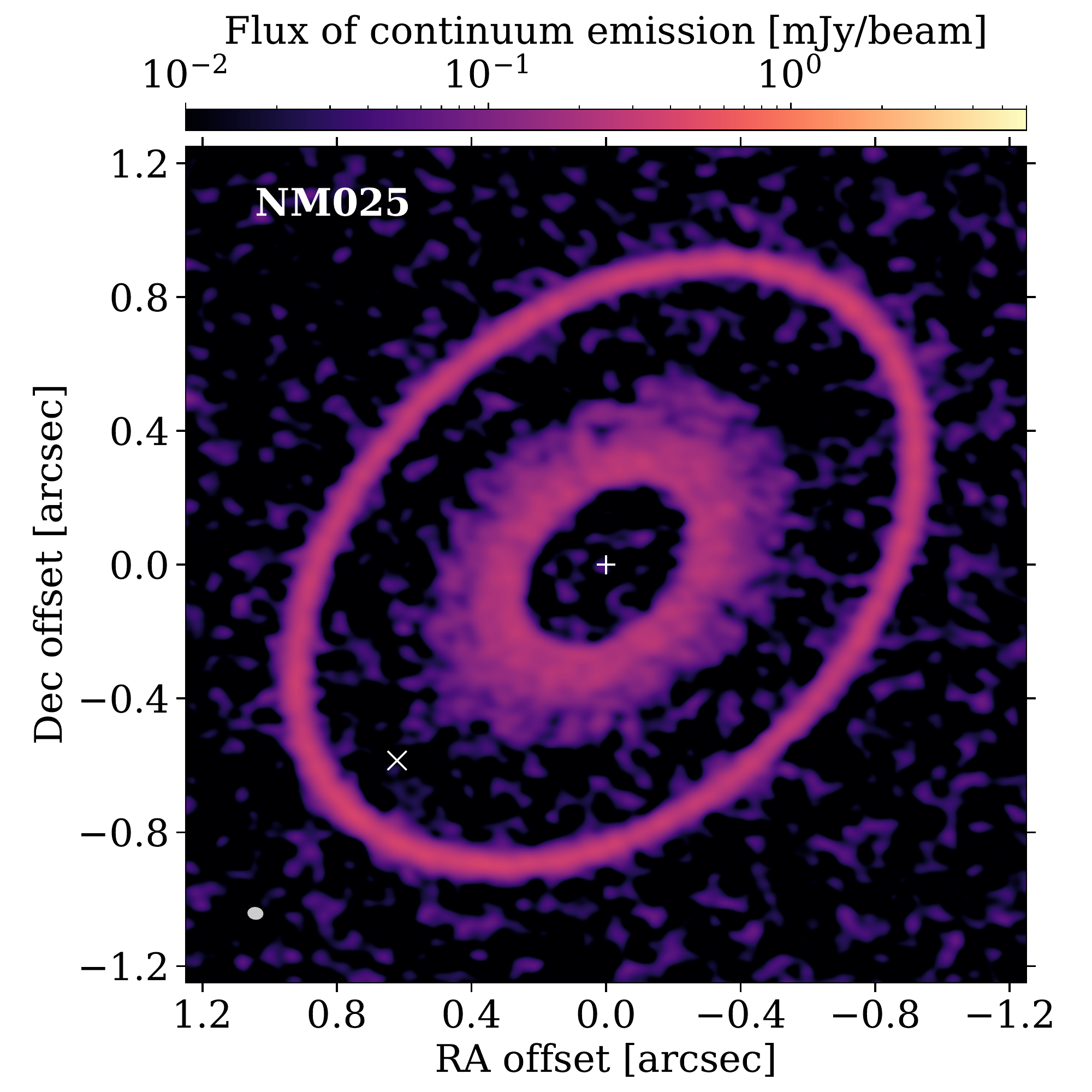}
\caption{\label{fig:fixed86au}Observed Band 6 continuum image (left-upper) of the HD 163296 disc obtained from the fiducial images data on the DSHARP data release \href{https://bulk.cv.nrao.edu/almadata/lp/DSHARP/}{website}, compared with the synthetic maps at the same wavelength obtained for our 3 models with a planet fixed at 86~au (the model's name is shown in the left-upper corner of each figure). The white â$\times$â symbol represents the position of the planet. The beam size is shown as a grey ellipse in the left-lower corner of each figure.}
\end{figure*}

For detailed analysis, we present results at 400 orbits at $R_{0}$ ($\sim 2.8 \times 10^{5}$ years with $R_{0} = 100$~au). This corresponds to just about 500 orbits at the planet's orbital radius at 86 au. We find that at this time in the simulation the width and depth of the gap in the gas reach a quasi steady-state\footnote{Some studies, e.g., \citet{Kanagawa16}, find that much longer integration times are needed to reach steady state if the disc viscosity is as low as $\alpha = 10^{-4}$. However, here we find that for such low viscosities synthetic images show significant deviations from the observed one already at early times. Evolving the simulations for longer would only increase the disagreement between synthetic and observed images. Finally, we find that the system evolves after the planet is allowed to migrate on timescales as short as ${\cal}O(100)$ orbits, which is much shorter than the relaxation time we use. To check this conclusion we extended the low viscosity simulation to more orbits, and found no change in our qualitative results. The only change is the amount of dust around L4, L5 points is smaller due to dust diffusion \citep[e.g.,][]{Dsharp7, wu2023wind}. Finally, the low viscosity discs are not the focus of our study, but the fact that low viscosity can generate different substructures in disc will help the understanding of our results.}. We then performed dust radiation transfer calculations and computed synthetic images of the dust continuum emission at 1.25 mm. The side-by-side comparison between our synthetic images and the DSHARP band 6 continuum data for HD 163296 is presented in Fig. \ref{fig:fixed86au}. Note that our radiative transfer calculations use a 3D spherical grid that extends the 2D polar grid of our hydrodynamical simulations. The grid is therefore truncated in the radial direction, which is why our synthetic maps feature an inner hole about the central star (it is thus not due to a real lack of dust material in our simulations).

Model NM05a4, which is shown in the lower left panel of Fig. \ref{fig:fixed86au}, appears to be the least promising fit to the observed emission map, since there is a pronounced vortex in the synthetic data. This large-scale vortex, which is located at the outer edge of the planet gap, forms due to the Rossby-Wave Instability setting in at that location \citep[e.g.,][]{rwi-2001}, and corresponds to the outermost asymmetric bright ring in the synthetic image. The low viscosity also leads to some dust still corotating with the planet at the time in the simulation, which is not seen in the observed image.

The synthetic image of Model NM025 is shown in the lower right panel of Fig. \ref{fig:fixed86au}. For this model we use $\alpha = 10^{-3}$ and $M_{\rm disc} = 0.025\,\msun$. No vortices appear in this case, but the gap carved by the planet in this model is much wider than the one seen in the observation. We also note that the disc region interior to the planet is dimmer than in the other models. The reason is two-fold. First, a lower disc mass implies a lower dust mass for a fixed dust-to-gas mass ratio. This contributes to lowering the overall dust emission, which turns out to be mostly optically thin in this model. Also, a lower disc mass implies faster dust drift, in particular for the 1 mm-sized dust which is therefore more depleted in the inner disc parts in Model NM025 compared to Model NM05 (this particular point will be illustrated in the left panels in Fig.~\ref{fig:dust_drift} for Model PM025, which includes planet migration).

The upper right panel of Fig. \ref{fig:fixed86au} shows the synthetic image obtained with Model NM05 for $\alpha = 10^{-3}$ and $M_{\rm disc} = 0.05\,\msun$. This model provides a closer match to the observations than previous Models NM025 and NM05a4. The synthetic map of Model NM05 does not show asymmetric structures as in the lower viscosity model, and its gap width and the mean dust continuum flux are more consistent with the observation.

Concluding this section, we see that with a fixed planet model it is possible to obtain a reasonable synthetic match to the data by adjusting the disc mass and its viscosity parameter. For the problem at hand, $\alpha = 10^{-3}$ and $M_{\rm disc} = 0.05\,\msun$ provide a decent agreement with the data.

\subsection{Planetary migration: starting point at 86~au}\label{sec:migration-86au}
In Section~\ref{sec:simovserve-fixed} we discarded planet migration, assuming that the planet orbit was fixed. However, given that some of our models feature a rather large disc mass, one may worry that a substantial migration of the planet could occur over the duration of the simulations. In addition, the planet ability to carve an annular gap in the disc, the dust dynamics and its emission could all be affected by planetary migration as well \cite[e.g.,][]{Meru19-Ring,TypeIII-Wafflard-Fernandez2020}.  Therefore, in this section we relax the fixed orbit assumption, letting the planet interact with the disc and hence migrate freely as directed by the gravitational torque from the disc (cf. Section~\ref{sec:planet}). 

For a migrating planet, its starting position may be defined as the orbital radius where the planet has reached its current mass. This radius is, however, very uncertain as the onset of runaway gas accretion and its pace depend on the physical properties of the disc in the planet's circumplanetary environment, in particular on its cooling rate, and on its heating rate via solid phase accretion \citep[e.g.,][]{AyliffeBate09,Szulagyi14,Bern20-2}. A variety of recent observational data and simulations actually indicate that planets may grow much slower than usually believed \citep[e.g., see the references and discussion in][]{nayakshin2022}. Therefore, we have adopted two different strategies about the starting position of the planet ($R_{\rm p, start}$): either it is released at 86 au, the same location as in Section~\ref{sec:simovserve-fixed}, which is what we assume in the current section, or it is released at an arbitrarily larger separation of 130 au, which is what will be assumed from the next section onward. We also define the planet's migration timescale, $t_{\rm m}$, as 
\begin{equation}
    t_{\rm m} = t_{\rm sim}\; \frac{R_{\rm p, start}}{R_{\rm p, start} - R_{\rm p, end}}, 
    \label{migration-timescale}
\end{equation}
where $R_{\rm p, end}$ denotes the planet's orbital radius reached at the end of the simulation, $t_{\rm sim}$. Table \ref{table1} lists $t_{\rm m}$ obtained in our simulations. 

\begin{figure}
\begin{centering}
\includegraphics[width=0.49\textwidth]{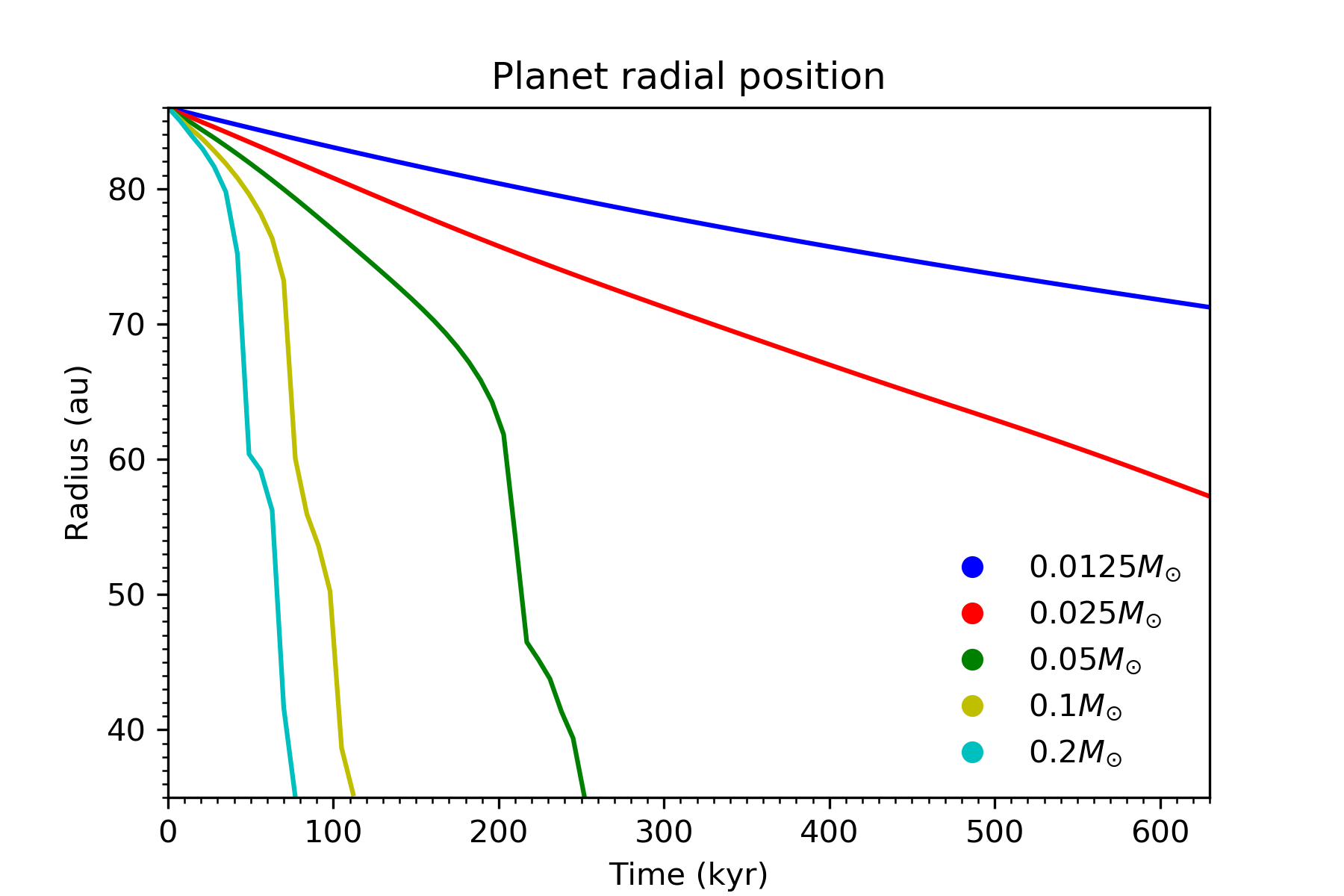}
\par\end{centering}
\caption{\label{fig:migration_rate}Time evolution of the planet's orbital radius in the simulations presented in Section~\ref{sec:migration-86au}, in which the planet starts migrating at 86 au after having remained on a fixed orbit over 100 orbital periods. Different curves correspond to different disc masses, the green curve (0.05 $M_{\odot}$) corresponds to the disc mass used in the best model described in Section~\ref{sec:simovserve-fixed}.}
\end{figure}

Since the rate of planetary migration depends sensitively on the disc's surface density, we have explored a range of disc masses from a minimum of $M_{\rm disc} = 0.0125\,\msun$ to a maximum value of $M_{\rm disc} = 0.2\,\msun$ in steps by a factor of 2. We display the time evolution of the planet's orbital radius in Fig. \ref{fig:migration_rate}. The maximum duration of our simulations, $t_{\rm sim, max}$, is approximately 630 kyr, which is about 1130 orbital periods at the planet's initial separation of 86 au. Some simulations are terminated earlier, however, if the planet reaches the inner boundary earlier than 630 kyr.

Fig.~\ref{fig:migration_rate} shows that, not surprisingly, the planet migration rate increases with disc mass \citep[see, e.g., the review by][]{BaruteauEtal14a}. So long as $M_{\rm disc} \leq 0.025\,\msun$, our planet migrates inward smoothly in a near type I migration fashion (the planet produces just a weak depression in gas around its orbit, as can be appreciated from the gas surface densities shown in Fig. \ref{fig:dust_drift}). For $M_{\rm disc} \ge 0.05\,\msun$, however, our planet migrates inward at a highly variable rate, with several stages of fast, accelerated type III (runaway) migration, sandwiched between stages of slower migration. Intermittent runaway migration has already been observed and documented in several studies \citep[e.g.,][]{ TypeIII-Lin2010, TypeIII-McNally2019, TypeIII-Wafflard-Fernandez2020}. The intermittency of inward runaway migration is intimately related to the time evolution of the planet's vorticity-weighted coorbital mass deficit (for definition see \citealp{TypeIII-Masset2003}), which can fluctuate around the planet mass depending on the surface density profile of the background gas, and how strongly the inner wake of the planet shocks the background gas \citep{TypeIII-Wafflard-Fernandez2020}. As can be seen with the green curve in Fig. \ref{fig:migration_rate}, fast, intermittent inward runaway migration is obtained indeed for $M_{\rm disc} = 0.05~\msun$, which was the disc mass selected for our ``best-fit'' model in Section~\ref{sec:simovserve-fixed}.

As argued by \cite{nayakshin2022}, {\em very rapid} planet migration is a challenge to the widely accepted scenario in which {\em all} of the gaps/rings seen by ALMA are due to embedded planets \citep[e.g.,][]{LongEtal18}. In regards to HD 163296 in particular, we find that for $M_{\rm disc} \ge 0.05~\msun$ the migration time as defined by Eq.~(\ref{migration-timescale}) is shorter than $\sim 0.4$~Myr. This is very short compared with the estimated age of this system ($t_* \sim$ 7 to 10 Myr). Another way to look at this is to define the gap crossing time $t_{\rm gap}$, which is the time the planet spends while migrating over the full radial width of its annular gap. This time is shorter than $t_{\rm m}$ by approximately $\Delta R/R \approx 5\times(q/3)^{1/3}$, since the typical radial width of a planetary gap (in the gas) carved by a giant planet is about 5 Hill radii \citep{Masset2006}. Thus, to stand a decent chance of observing a giant planet passing through the 86 au gap and also opening it, HD 163296 needs to be hatching gas giant planets beyond 86 au at a rate of approximately $t_{\rm gap}^{-1}$. One can then estimate the number of gas giant planets that HD 163296 needs to produce over its lifetime as 
\begin{equation}
    N_{\rm p} \sim \frac{t_*}{t_{\rm gap}}\sim 100
\end{equation}
for $q=1.3\times 10^{-4}$, $t_* = 7$ Myr and $t_{\rm m} = 0.4$ Myr. This is probably excessive. More generally, it is unlikely that a planet is in the gap at the time of observation if it is migrating rapidly, unless there are $N_{\rm p}$ planets passing through the gap in succession. On these grounds we conclude that planet migration constrains the disc mass in HD 163296 to be no larger than $\sim 0.025~\msun$.

\subsection{Planetary migration: starting point at 130~au}\label{sec:migration-130au}
Previous Section~\ref{sec:migration-86au} shows that there ought to be an upper limit for the mass of the disc around HD 163296 to avoid fast inward migration of the putative planet responsible for the dark ring of continuum emission near 86 au. The simulations presented in this section assume that this putative planet formed and reached its final mass further out in the disc, specifically at 130 au. The planet is introduced by increasing its mass to its final value over 10 local orbital periods during which the planet remains on a fixed orbit. Simulations then let the planet migrate in the disc, and are terminated when the planet reaches 86 au. In this section we explore two disc masses: 0.025 $\msun$ (Model PM025) and 0.05 $\msun$ (Model PM05). For both disc masses, the migration from 130 au down to 86 au proceeded smoothly, without runaway, and lasted for about 0.68 Myr and 0.30 Myr for models PM025 and PM05, respectively. For completeness, we have checked by restarting both of these simulations that, below 86~au, the planet qualitatively has a similar orbital evolution as in the simulations where the planet was inserted at 86~au. In particular, the planet in Model PM05 undergoes a stage of accelerated type III migration between $\sim$65 and $\sim$45~au, much like in Model OM05. This indicates that the exact starting position of the planet (130 au here) is not very important so long as one runs the simulations long enough so that the planet eventually reaches its current ``observed" position at 86 au.

\begin{figure*}
\centering
\includegraphics[width=0.49\hsize]{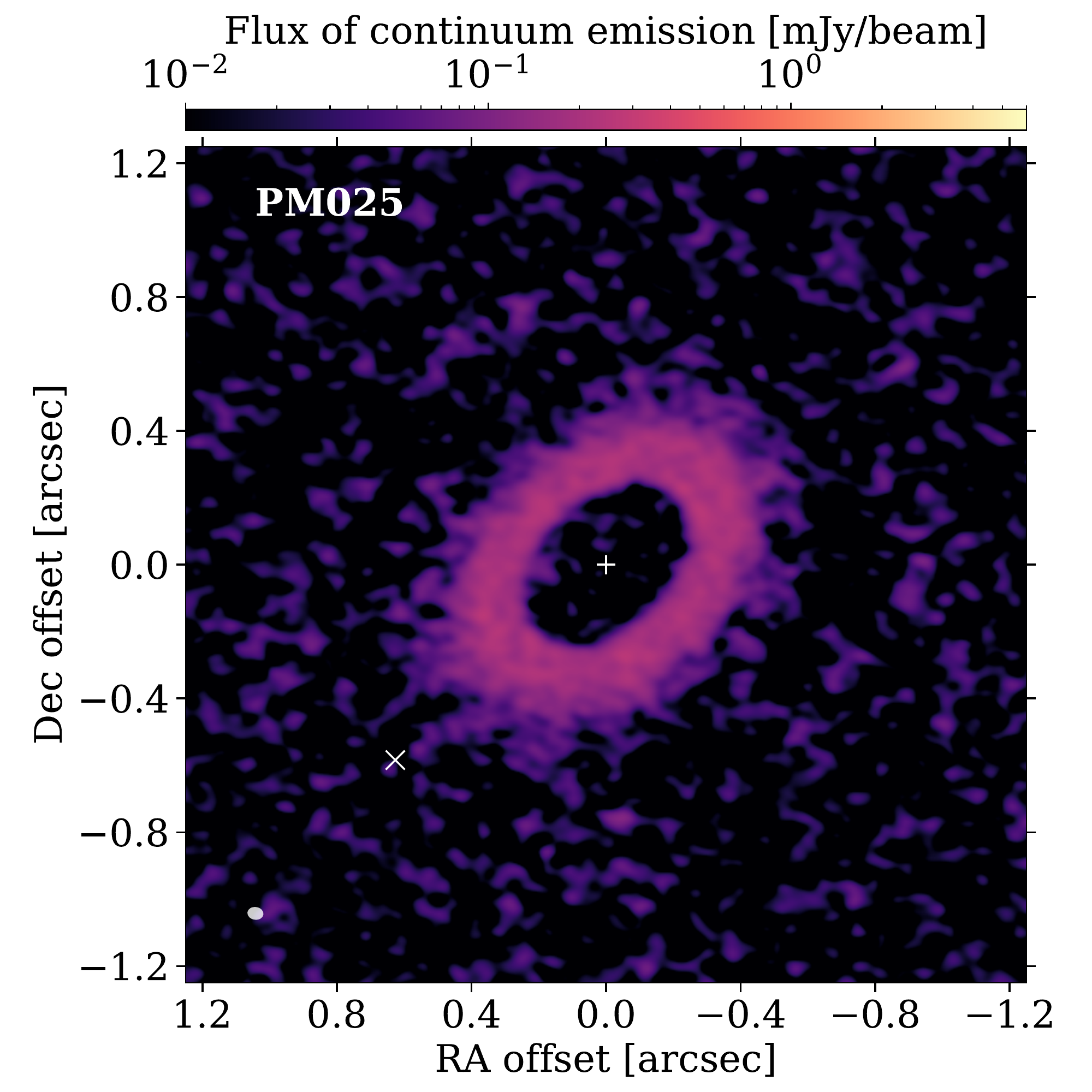}
\includegraphics[width=0.49\hsize]{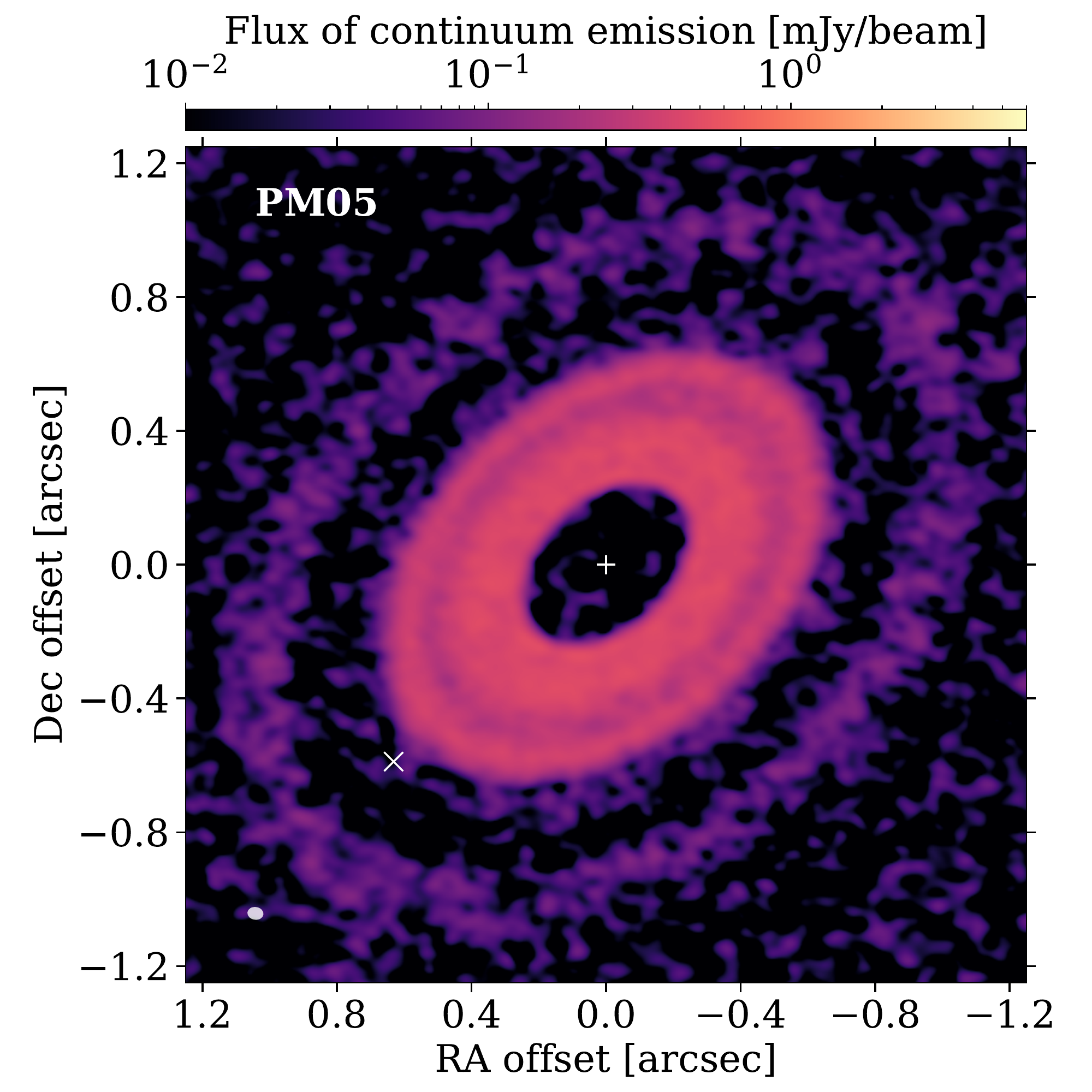}
\caption{\label{fig:migration_130}Similar to Fig. \ref{fig:fixed86au}'s synthetic maps but for Model PM025 (left) and Model PM05 (right).}
\end{figure*}

Fig.~\ref{fig:migration_130} shows the synthetic map of continuum emission for both models when the planet reaches about 86 au. The same colour scale as in Fig.~\ref{fig:fixed86au} is used, to provide a better comparison with the observed map. For Model PM025, shown on the left-hand side, we no longer see a bright ring of emission outward of the planet gap compared to the fixed planet model at same disc mass (Model NM025) displayed in the lower-right panel of Fig.~\ref{fig:fixed86au}. We only see a dim emission inward of the planet gap at a similar level of flux as in Model NM025. A similar comparison can be made between Model PM05, shown on the right-hand side of Fig.~\ref{fig:migration_130}, and Model NM05, shown in the top-right panel in Fig.~\ref{fig:fixed86au}: the outer bright ring is much dimmer with Model PM05, and we also notice a secondary dark ring inward of the planet gap at an orbital separation $\sim$60~au. Since the dust temperature is the same in all our synthetic maps (recall that it is simply the gas temperature in the hydrodynamical simulations), the overall flux differences between Models PM025 and PM05 therefore indicate that the dust's continuum emission is mostly optically thin for both models.

\begin{figure*}
\centering
\includegraphics[width=0.49\hsize]{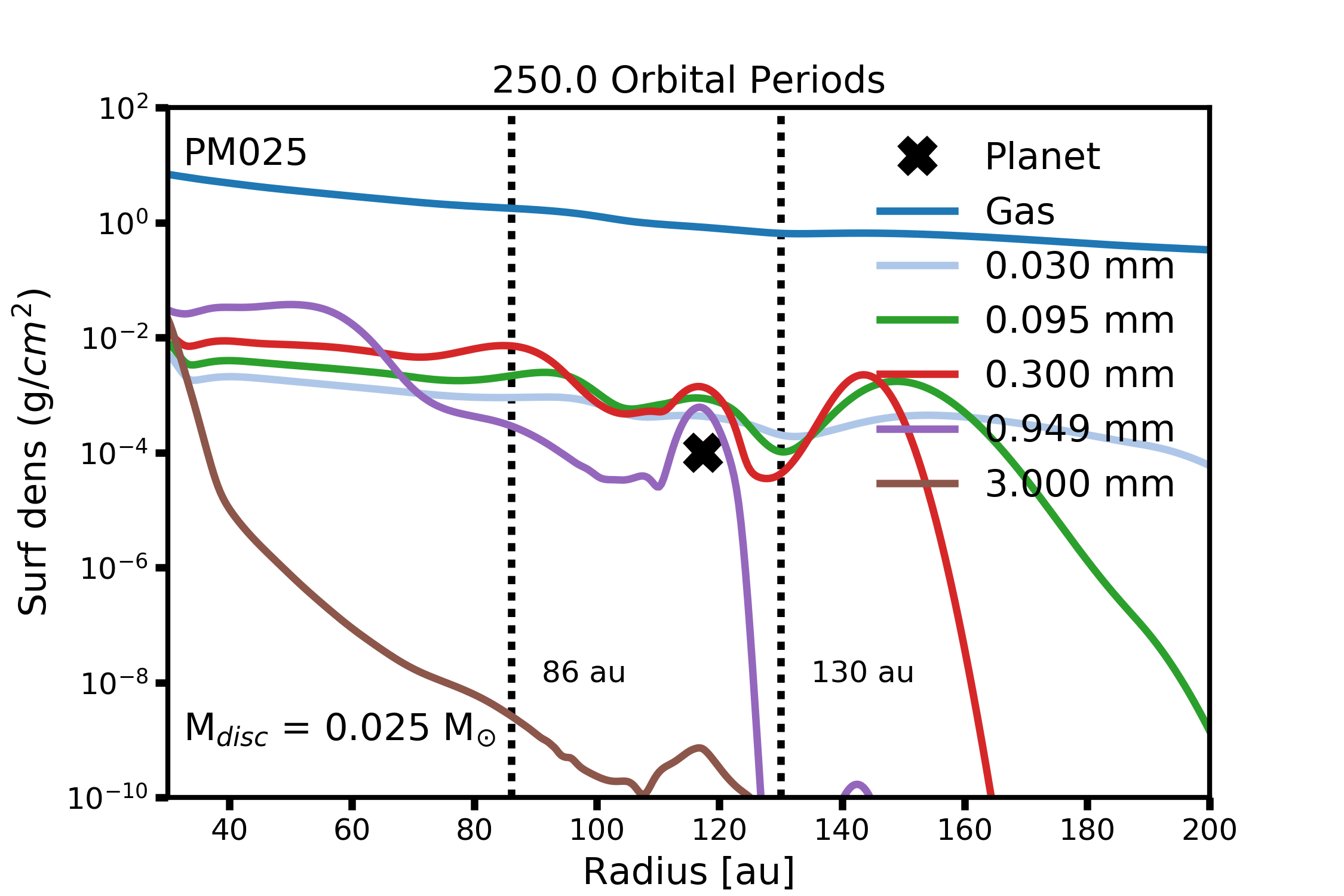}
\includegraphics[width=0.49\hsize]{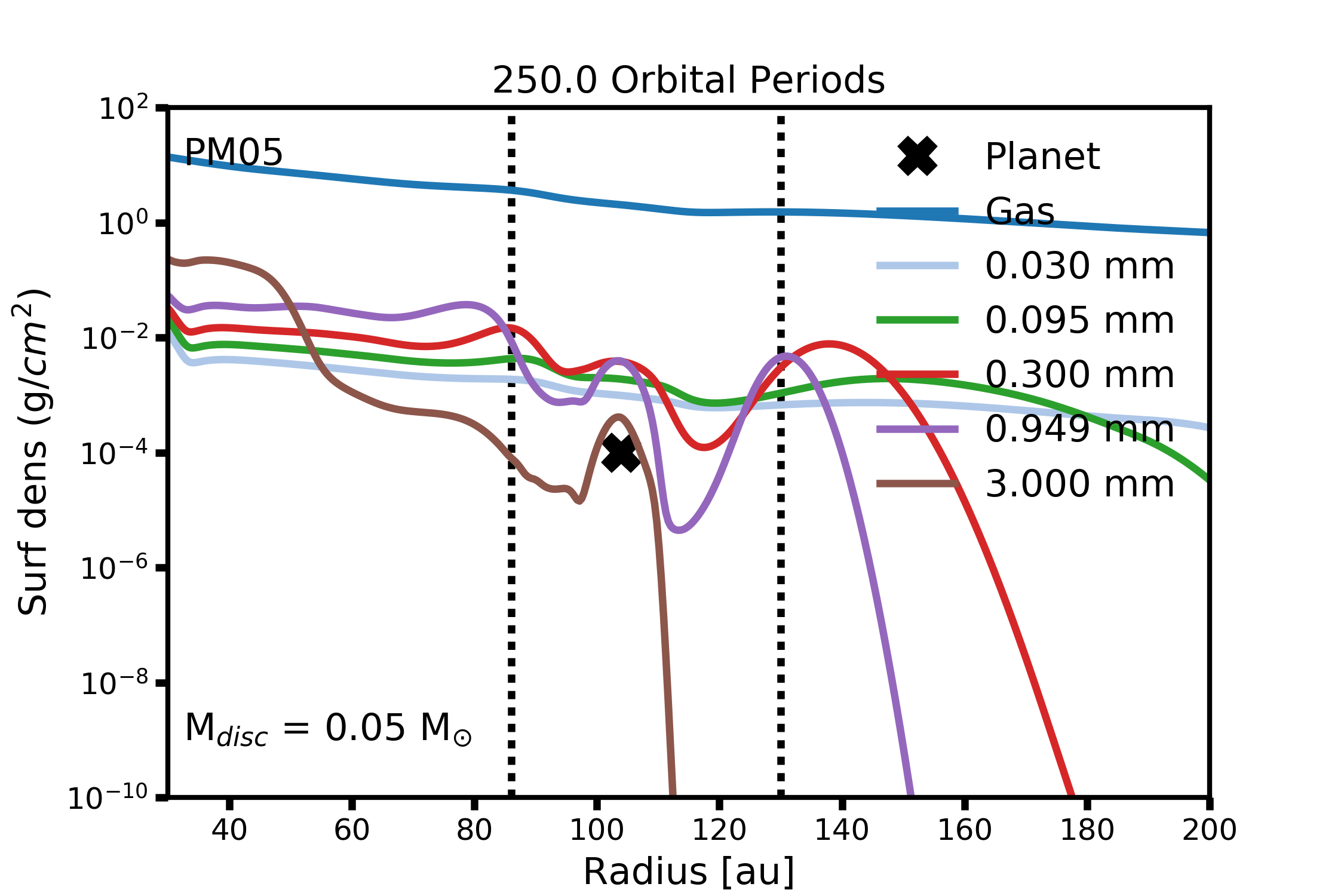}
\includegraphics[width=0.49\hsize]{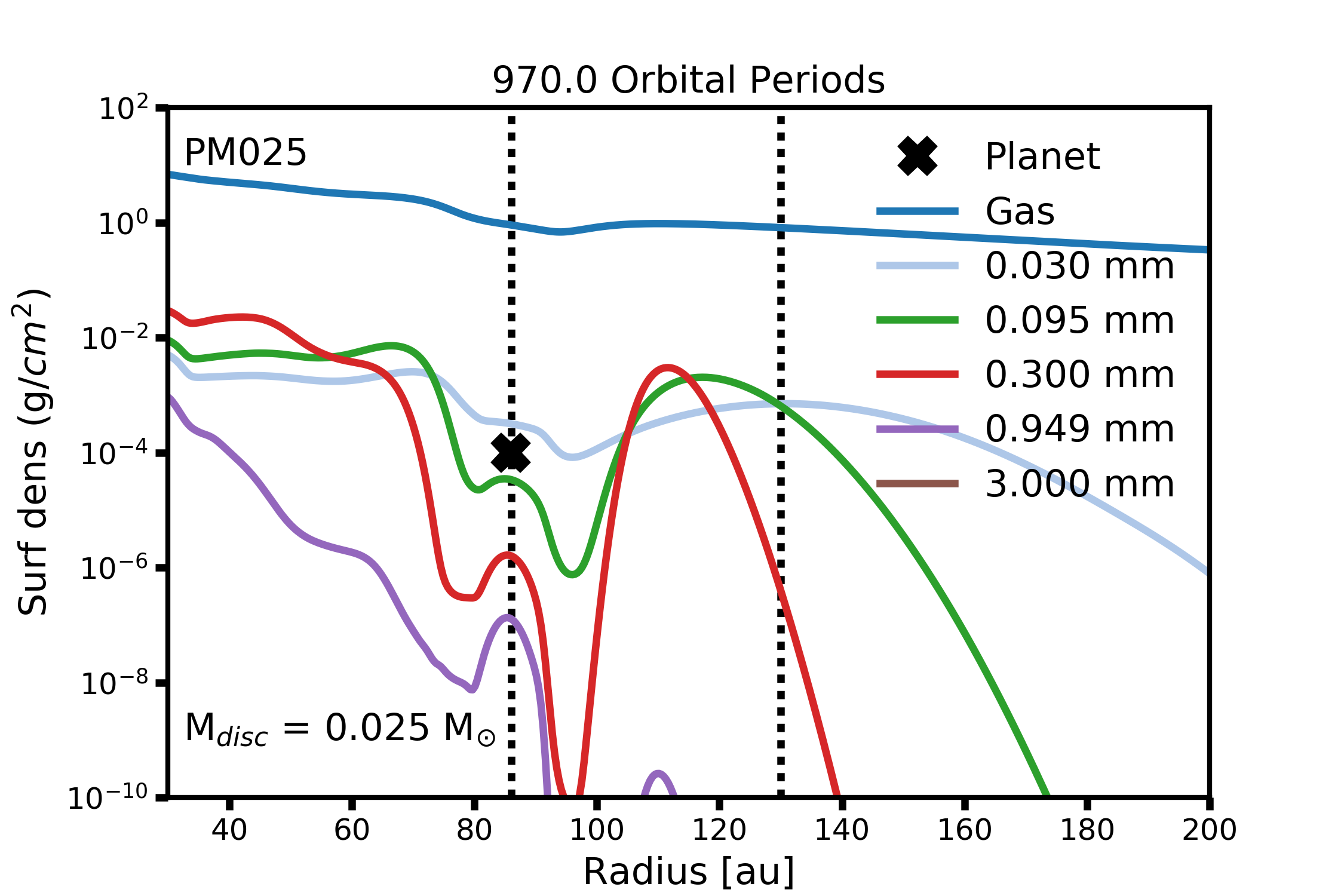}
\includegraphics[width=0.49\hsize]{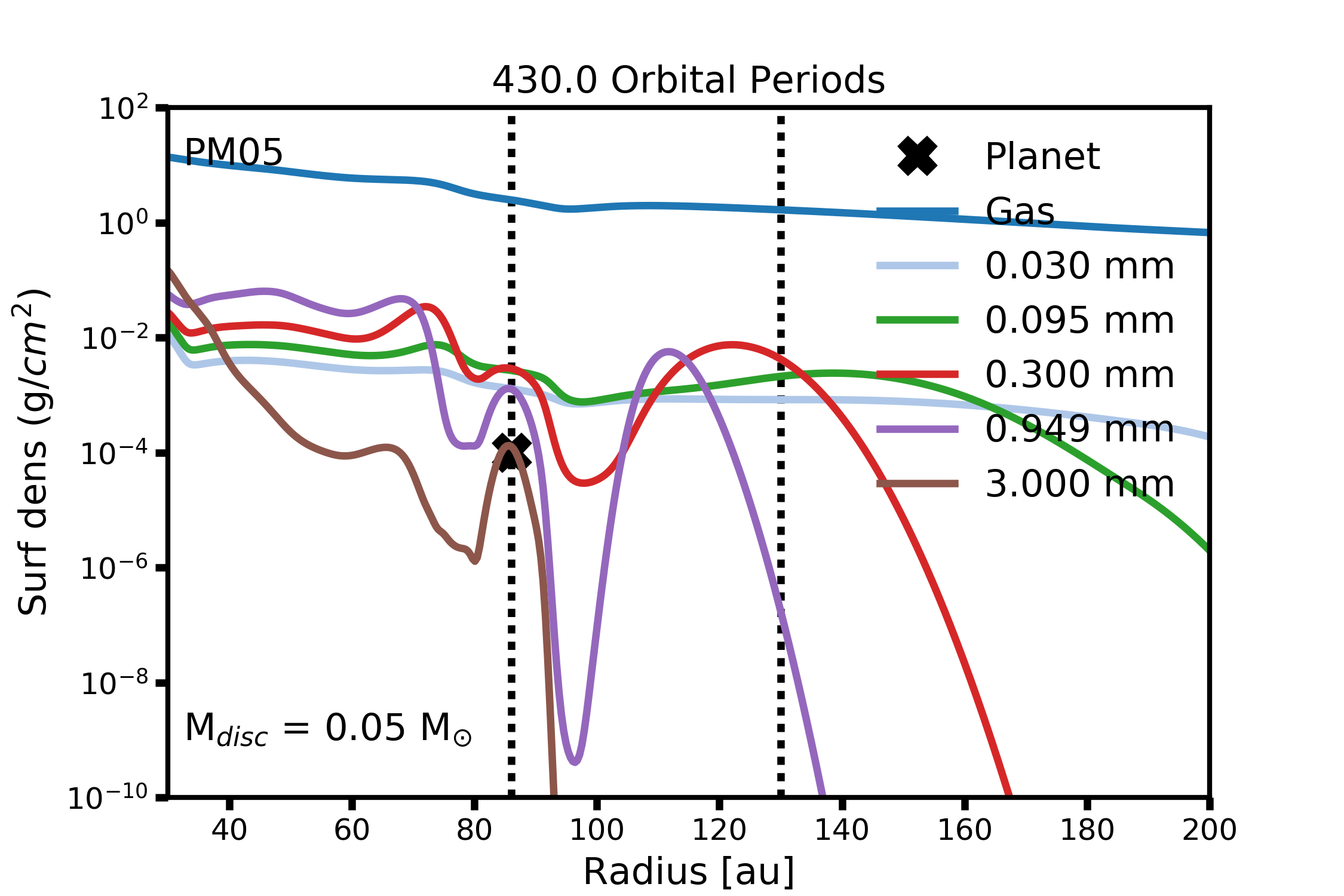}
\caption{\label{fig:dust_drift}Surface densities (in g cm$^{-2}$) of gas and dust fluids in Model PM025 (left panels) and Model PM05 (right panels) resulting from our hydrodynamical simulations at two different times: at 250 orbits (upper panels) and when the planet reaches 86 au (lower panels). Different colours correspond to different dust sizes (see legend in each figure). The black cross symbol marks the position of the planet at each time in the simulation. The vertical dotted lines show the initial (130 au) and final (86 au) positions of the planet in the simulations.}
\end{figure*}

To get more insight into the differences in continuum emission between the fixed and migrating planet models, we display in Fig.~\ref{fig:dust_drift} the  azimuthally-averaged surface density profiles of the gas and of each dust fluid for Models PM025 (left panels) and PM05 (right panels). Profiles are shown at two epochs: at 250 orbital periods at 100 au (0.17 Myr) in the upper panels, and when the planet has reached $\sim86$ au. The panels illustrate how dust radial drift and dust trapping about the planet vary with dust size and disc mass. 

For Model PM025, we see that both 1 mm and 3 mm dust fluids drift inward faster than the planet migrates in. This, and the smaller dust mass, explain why the continuum emission at 1.3 mm from the inner disc is dim in this model (as already discussed in Sect.~\ref{sec:simovserve-fixed}). 

Concerning the outer bright ring, the fact that the planet migrates in Model PM025 relatively rapidly implies that it cannot trap the largest dust outside its annular gap as vigorously as in Model NM025. This explains the lack of a strong bright ring outward of the planet gap in Model PM025 compared with NM025.

Now, for model PM05, only the 3 mm dust fluid tends to drift inward faster than the planet migrates in, while the 1 mm dust drifts at a pace that is no faster than the planet's migration rate. The presence of the 1 mm dust in the inner disc regions as well as beyond the planet gap explains the different flux distributions of Models PM025 and PM05 \citep{Meru19-Ring, Nazari+2019}. Still, the lack of 3 mm dust trapped at the outer edge of the planet gap in Model PM05 explains why the outer bright ring in PM05 is dimmer compared to Model NM05, where the fixed planetary orbit helps to build up  a stronger pressure maximum outside of the planet location. This can be further appreciated by inspection of the dust surface density profiles for Model NM05, which are shown in the left panel of Fig.~\ref{fig:dust_drift_porous}. We finally point out that the dark ring of emission  just inward of the planet location in Model PM05 is mainly related to the local minimum in the dust surface density of the 3 mm dust fluid, which can be seen around 60 au in the bottom-right panel of Fig.~\ref{fig:dust_drift}.

\subsection{Dust material density: compact or porous?}\label{sec:dust-result}
Previous section shows that the smaller the mass of the disc in HD 163296, the longer the migration timescale for the putative planet at 86~au. This relieves the statistical requirement for HD 163296 disc to hatch dozens of planets (cf. the end of Section \ref{sec:migration-86au}) but then we encounter the dust drift challenge. The smaller the disc mass, the faster is dust radial drift,  and hence the smaller the flux of continuum emission compared to observations. We find ourselves at an impasse. At high disc mass the planet is lost into the star too quickly, at low disc masses the dust drifts in too rapidly; and there appears no intermediate value of disc mass that would resolve these opposite challenges.

To avoid such a fast radial drift, we explore in this section the possibility that dust particles have a smaller material density than conventionally assumed. More specifically, we now set the internal density of dust particles to $\rho_{\rm d}=0.1~{\rm g\;cm}^{-3}$, like in \citet{fargo2radmc3d-dust}, and we refer to them as porous dust particles (as opposed to compact dust particles for our default internal density). The basic idea is that porous dust particles will have lower Stokes numbers than compact particles at a given disc mass (Eq.~\ref{stokes_number}), and therefore a slower radial drift. Their absorption opacity will differ too, and dust radiative transfer calculations are needed to check how the overall level of flux emission is impacted.

\begin{figure*}
\centering
\includegraphics[width=0.49\hsize]{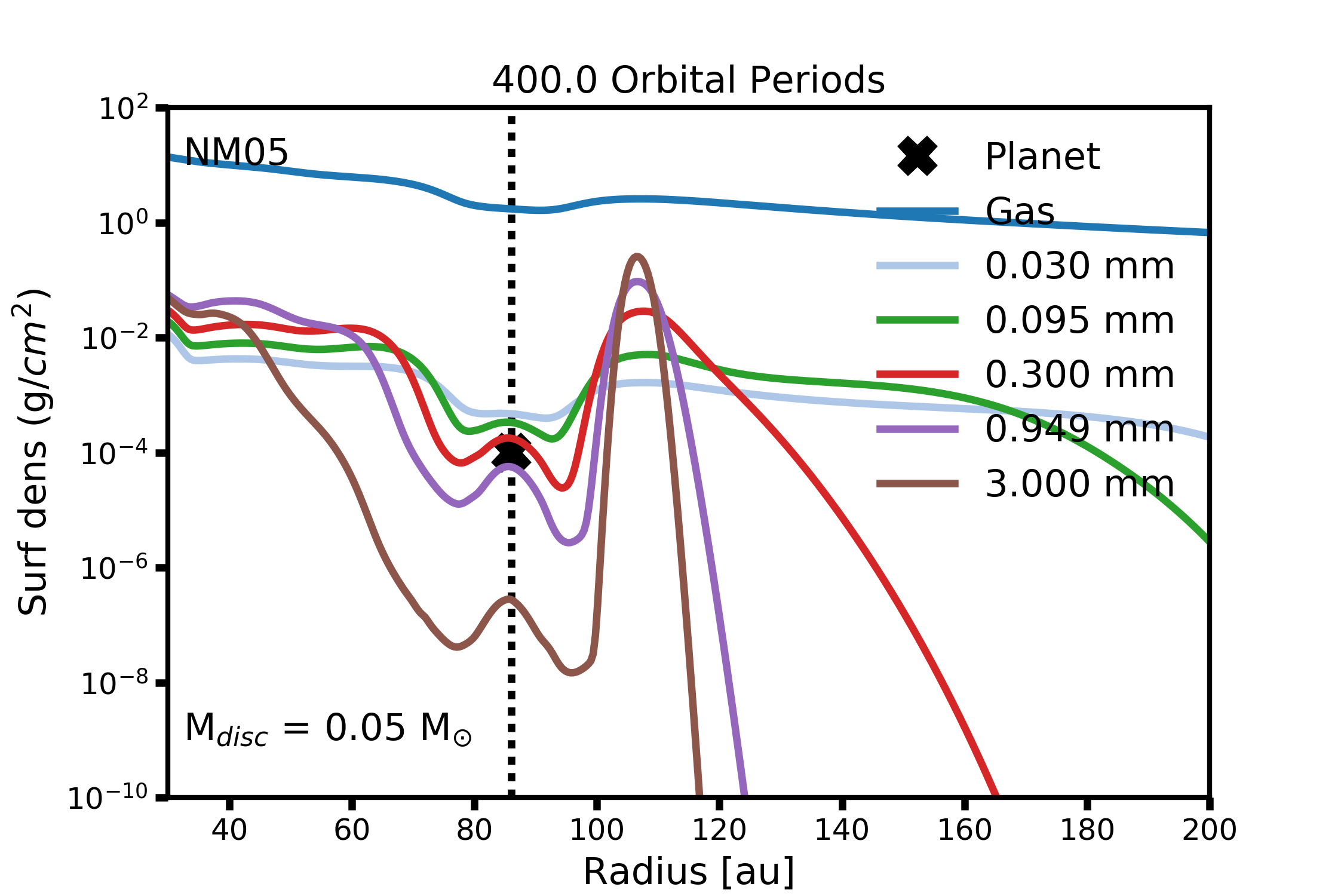}
\includegraphics[width=0.49\hsize]{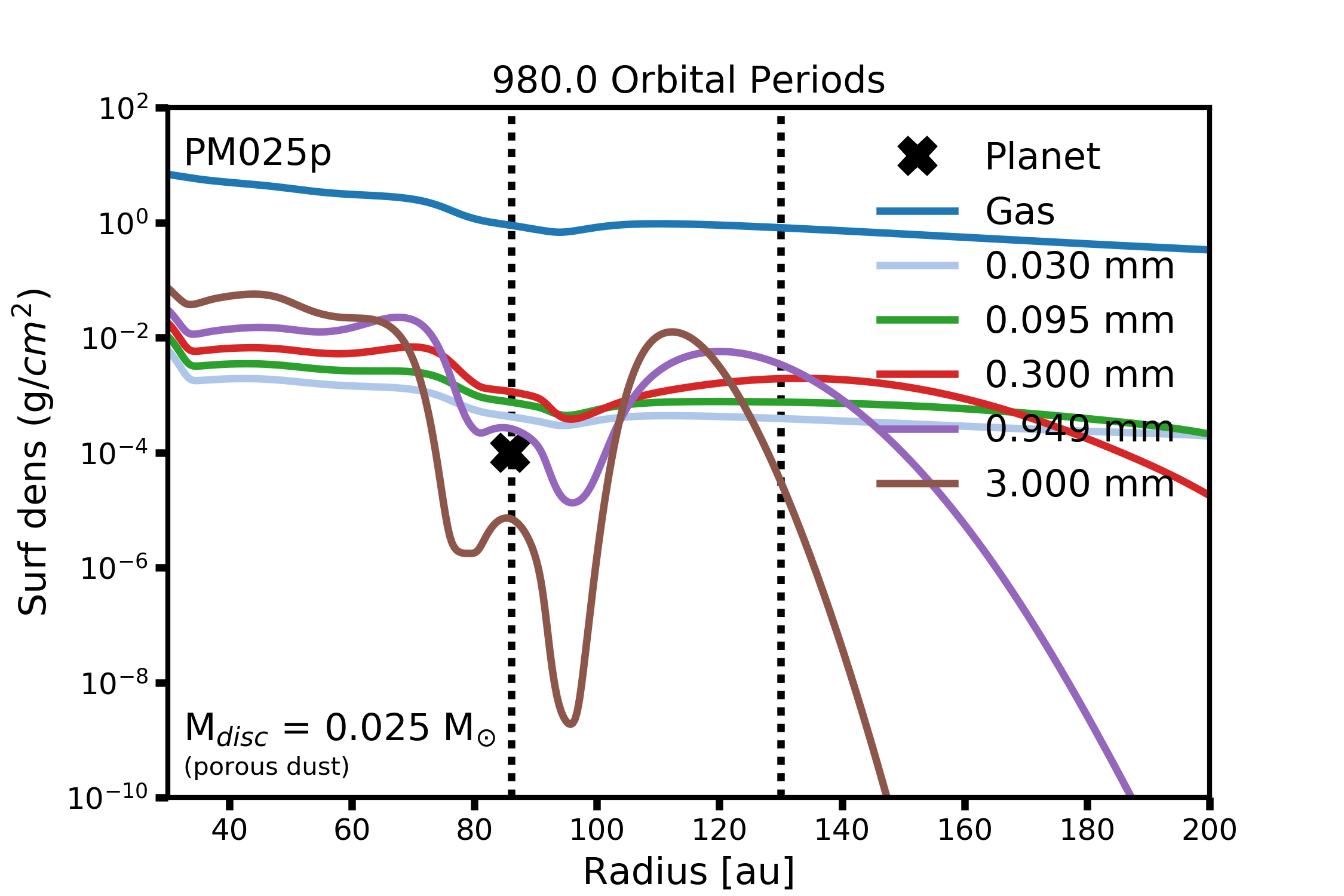}
\caption{\label{fig:dust_drift_porous}Similar to Fig. \ref{fig:dust_drift} but for Model NM05 (left) and Model PM025p (right). For NM05, surface density profiles are shown at the same time as  in the synthetic image in the top-right panel in Fig.~\ref{fig:fixed86au}. For PM025p, the time is that when the planet reaches 86~au.}
\end{figure*}

\begin{figure*}
\centering
\includegraphics[width=0.49\hsize]{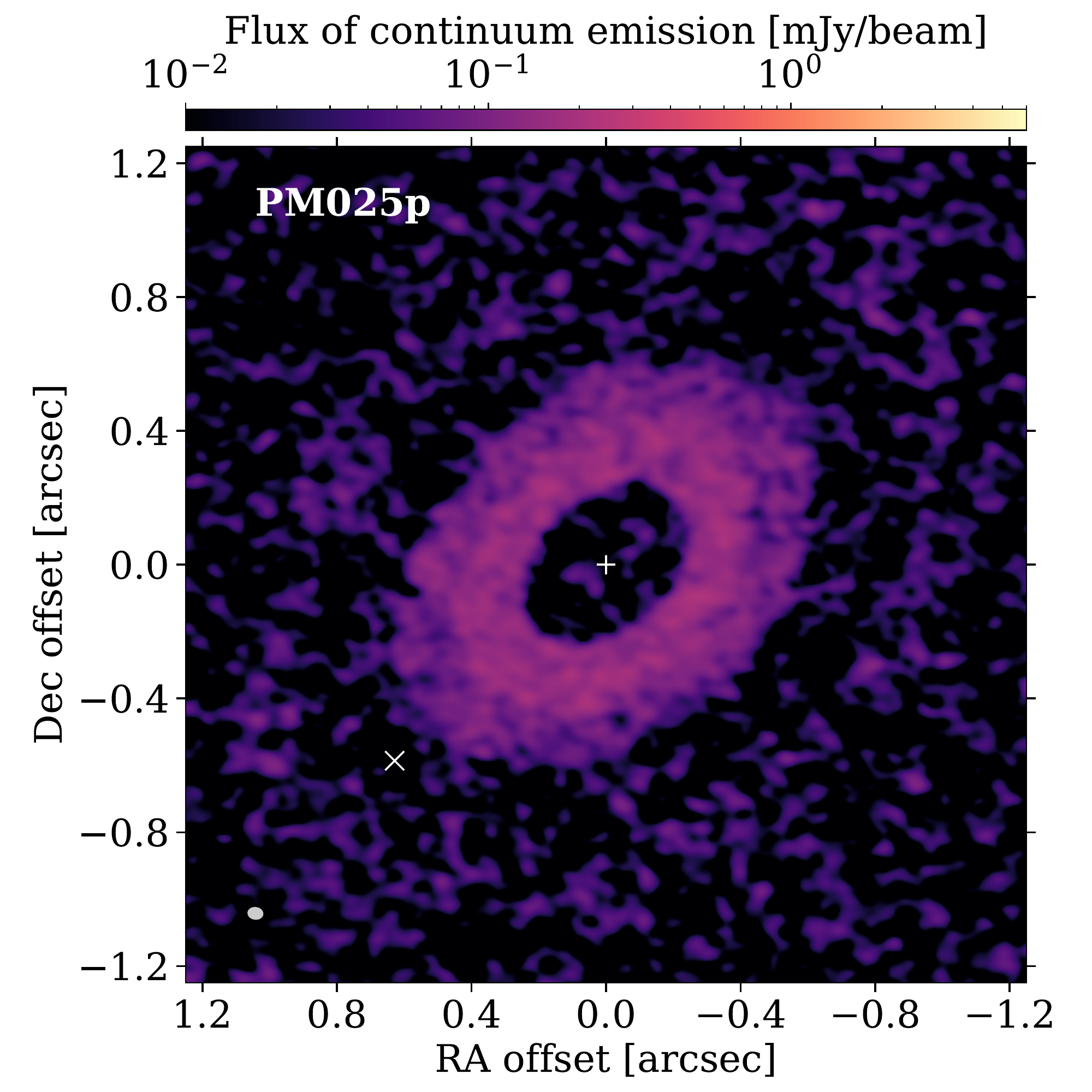}
\includegraphics[width=0.49\hsize]{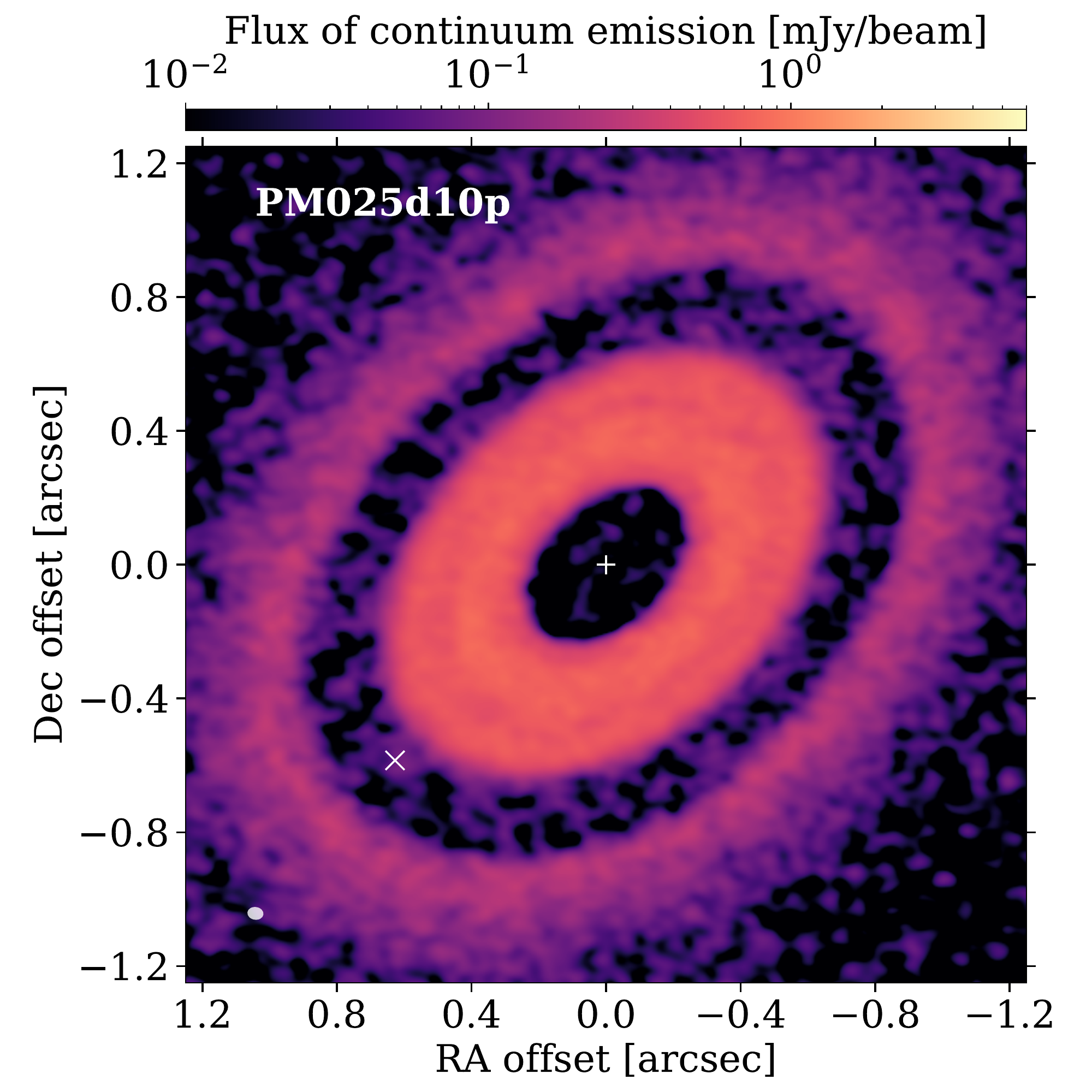}
\caption{\label{fig:porous_migration_130}Similar to Fig. \ref{fig:migration_130} but for Model PM025p (left) and Model PM025d10p (right).}
\end{figure*}

The right panel in Fig.~\ref{fig:dust_drift_porous} shows the gas and dust  surface density profiles for simulation PM025p, for which the disc mass is 0.025~$\msun$ and porous dust particles are adopted. The density profiles are those when the planet reaches about 86~au, and they can be directly compared to those obtained in the previous section for compact particles (bottom-left panel in Fig.~\ref{fig:dust_drift}). A comparison between both panels shows that, as expected, dust radial drift is much slower for porous particles. At the time when the planet reaches 86~au, the surface density of 1 and 3 mm dust particles is high and close to its initial value. On the other hand, the synthetic image displayed in the left-hand panel in Fig.~\ref{fig:porous_migration_130}, shows a very low level of continuum flux, which arises from porous dust particles having a smaller absorption opacity compared to compact particles of same size \citep{Kataoka14-porous-dust}.

We thus further tested increasing the dust-to-gas mass ratio $\epsilon$. As shown in the right-hand panel in Fig. \ref{fig:porous_migration_130}, increasing $\epsilon$ from 0.01 to 0.1 gives a satisfactory agreement between the properties of the synthetic map (overall level of flux, gap width etc.) and those of the ALMA observation.

\section{Concluding remarks}\label{sec:conclusion}
There is mounting observational evidence that annular sub-structures are very common features of discs continuum emission. Interestingly, not only are bright and dark rings frequently observed, like amongst the Taurus discs \citep{LongEtal18}, they also tend to come as multiple-ring systems \citep{Dsharp2}. This is the case of HD 163296 disc, for which the detection of localised velocity perturbations in the CO emission strongly hints at the presence of massive planets in (some of) its dark rings \citep{Teague18, Pinte20-Dsharp-Vkinks}. The wealth of observational data on the HD 163296 disc clearly makes it an ideal target to test predictions of disc-planet interactions. Though many numerical works have intended to reproduce observed annular structures with planets (a statement that is not restricted to HD 163296 disc), only a handful of them, including this work, have taken planetary migration into account.

\begin{figure}
\begin{centering}
\includegraphics[width=0.49\textwidth]{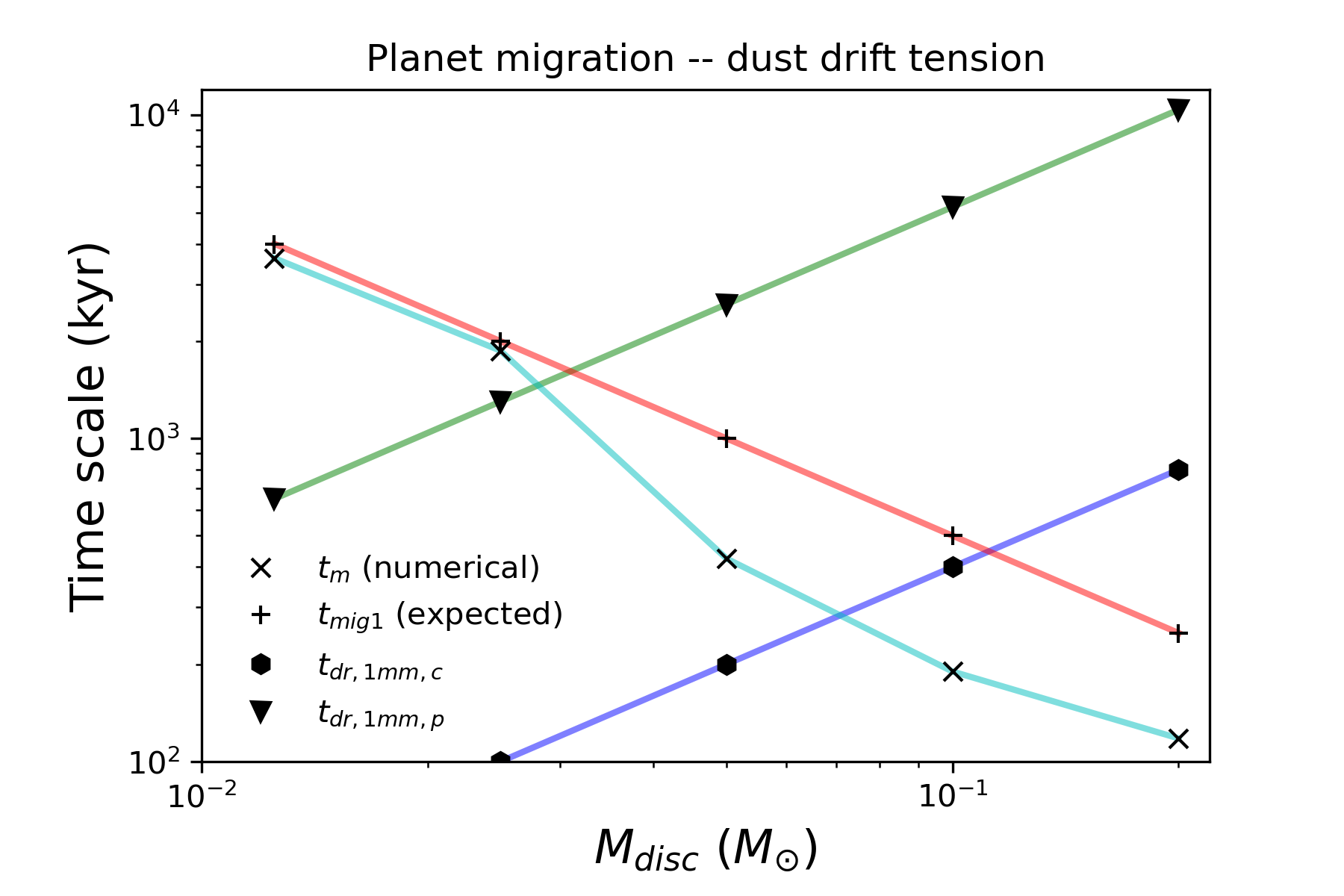}
\par\end{centering}
\caption{\label{fig:time_scale_planet_and_dust} Planet migration and dust drift time scales versus disc mass. In particular, $t_{\rm mig1}$ is the expected type I (eq. \ref{mig1-empirical}) and $t_{\rm m}$ is numerically determined migration timescales (eq. \ref{migration-timescale} and Table 1). The later is shorter due to the presence of runaway migration. The dust drift timescale $t_{\rm dr}$ are given by eq.~\ref{t_drift} for 1~mm dust particles for either compact (bottom blue line) or porous (top green) dust models. For compact dust there is no acceptable solution for the disc mass: at low disc masses the dust drifts into the star too quickly, whereas at high disc masses the planet is lost into the star by runaway migration. Porous dust model largely alleviates these issues (see text in Section 4).}
\end{figure}

Taking the HD 163296 disc as a case study, here we have shown that planetary migration could be used to set an upper limit on the mass of the disc. For the HD 163296 disc this upper limit is $\sim$ 0.025 $\msun$, which is at the low end of values compatible with the gas or dust emission, this disc mass is also lower than previous recommended values \citep[e.g., $\sim 0.15 M_{\odot}$, see][]{MAPS-Zhang}. While a lower disc mass would result in slower migration, and therefore to a higher probability of catching a migrating planet sculpting its disc, it would also lead to a faster dust radial drift. As a result we found no value for a disc mass that is compatible with both planet migration and dust emission constraints. The situation is illustrated in Fig.~\ref{fig:time_scale_planet_and_dust}.

The gist of the argument can be glipsed from the following semi-analytic argument. The theory of type I migration in isothermal discs leads to migration timescales depending on disc properties as $t_{\rm mig1} = C h^2 M_*^2/(\Omega_{\rm K} M_{\rm p} \Sigma R^2)$, with dimensionless coefficient $C$ depending on local disc properties \citep[e.g.,][]{Tanaka02,Paardekooper2010}. The disc surface density  is proportional to the total disc mass $M_{\rm d}$ through Eq.~(\ref{sigma-disc0}), and therefore we see that this scales as $t_{\rm mig1} \propto 1/M_{\rm d}$. Using the results of our lower mass OM125 and OM025 simulations that showed no runaway migration to deduce the constant $C$ from simulations directly, we get
\begin{equation}
    t_{\rm mig1} \sim 2\times 10^6\;\hbox{years}\; \frac{0.025~\msun}{M_{\rm d}} \;\frac{0.26 \mj}{M_{\rm p}}\;.
    \label{mig1-empirical}
\end{equation}
This time scale is short but not overly so compared with the age of HD 163296. Our hydrodynamical simulations actually place a surprisingly robust upper limit on $M_{\rm d}$ that is much more stringent than can be expected based on type I migration theory (Eq.~\ref{mig1-empirical}). We find indeed that at disc masses higher than $\sim 0.025~\msun$ runaway migration \citep[e.g.,][]{BaruteauEtal14a} shortens planet migration time to a fraction of a Myr. Since the gap crossing time is yet shorter than planet migration timescale (see the end of Section 3.3), we then estimated that up to 100 planets need to be injected into the disc of this system to present a decent statistical chance of us seeing the planet.

At the same time, the dust particles drift timescale \citep[e.g.,][]{Armitage07} is also a function of disc mass, albeit one that increases with $M_{\rm d}$  in the limit of small Stokes numbers:
\begin{equation}
    {t_{\rm dr}} = \frac{1}{\Omega_{\rm K}}\frac{\St^{-1}}{h^{2}\eta} \sim  10^5\;\hbox{years}\; \frac{M_{\rm d}}{0.025\,\msun}\; \frac{1\text{ mm}}{s} \; \frac{1.3 \text{ g cm}^{-3}}{\rho_{\rm d}},
    \label{t_drift}
\end{equation}
where $\eta = -d\ln \langle P \rangle / d\ln R \approx 3.2$ with $\langle P \rangle$ the azimuthally-averaged disc pressure (this estimate is made at $R=86$ au).

Comparing Eqs.~(\ref{mig1-empirical}) and (\ref{t_drift}) we see that there is no value of disc mass that allows both gas and dust to be present in the disc of HD 163296 for a time comparable with the $7-10$~Myr old age of this system. This is clearly a significant challenge to our understanding of how this system evolved to be in the current state.

We find that a porous rather than compact dust composition presents a potential solution to this dust drift versus planet migration tension. Simulations with strongly decreased dust density, $\rho_{\rm d} = 0.1$ g cm$^{-3}$, have shown that the dust can in fact stay in the disc for far longer. This is natural given Eq.~(\ref{t_drift}). To account for the observed level of dust continuum flux in HD 163296, the disc needs to be rather dust-rich, with a dust-to-gas mass ratio about 5 to 10 times larger than the canonical value of 0.01 used in the literature. However, the total dust mass that we infer is about 0.002 $\msun$, which is actually in good agreement with that inferred from the dust continuum emission \citep{Powell19-DustLanes}. Such a high dust-to-gas mass ratio in an old disc would imply that there were significantly more gas in the disc at earlier times, and that most of it, say $\sim 90$\%, has since been removed from the disc. Such a process must selectively remove gas but not dust.

Gas could be lost from the disc via gas accretion onto the star. However, as we need dust particles to be small in order to minimise dust drift into the star, we would expect the dust to be carried into the star with the gas. It is hence difficult to see why gas accretion onto the star would selectively remove gas from the disc. On the other hand, there have been suggestions that dust in older protoplanetary discs may be of secondary nature \citep[e.g.,][]{Turrini-19-2nd-dust}. In this scenario primordial dust is locked into planetesimals relatively early on. Once giant planets form, planetesimals are excited onto eccentric orbits. Like in debris discs \citep[e.g.,][]{Wyatt08}, planetesimal collisions then give rise to fragmentation cascade leading to dust production. Secondary dust has in fact been proposed for HD 163296 \citep{DAngelo-22-2nd-dust}. Gas outflows from the disc can also remove gas preferrentially, leaving larger dust particles behind \citep{Hutchison15-dusty-dinds}.

As a final word, we recall that, for the sake of quantitative research, we have only studied the case of a single planet embedded in a protoplanetary disc with physical parameters inspired from the HD 163296 disc. Considering multiple planets could certainly impact their orbital evolution but it may not go against their natural tendency to undergo fast runaway migration if the disc is massive enough. As an example, we mention again the recent study of \citet{Juan2023-HD163296}, who have shown that the many substructures in the dust continuum emission at 1.25 mm wavelenth, including the intriguing comma-shaped feature inside the dark ring of emission near 54 au, could be reproduced by invoking four planets in a resonant chain. However, their model did not take into account planetary migration while the disc mass was taken to be $0.15\msun$. Should planet migration be accounted for in their simulation, planets would not end up in a resonant chain but be destabilised instead by fast encounters due to disc-induced runaway migration. This is shown in Sect.~\ref{sec:multi-planet} in the Appendix.

\section*{Acknowledgments}
We thank Juan Garrido-Deutelmoser for sharing his problem setting and the referee for helpful comments. We thank Yi-Xian Chen, Haochang Jiang, Kaitlin Kratter, Adrien Leleu and Douglas N. C. Lin for constructive suggestions. Y.W. further thanks Wenrui Xu, Shuo Huang, Shangjia Zhang and Kiyoaki Doi for helpful discussions. This research used the ALICE High Performance Computing Facility at the University of Leicester, and DiRAC Data Intensive service at Leicester, operated by the University of Leicester IT Services, which forms part of the STFC DiRAC HPC Facility (www.dirac.ac.uk). Part of the numerical simulations were performed on the CALMIP Supercomputing Centre of the University of Toulouse. Y.W. gratefully acknowledges the support by the DUSTBUSTERS RISE project (grant agreement number 823823) for his secondment in University of Arizona. S.N. acknowledges the funding from the UK Science and Technologies Facilities Council, grant No. ST/S000453/1.

\section*{Data availability}
The FARGO3D code is publicly available from \href{http://fargo.in2p3.fr/}{http://fargo.in2p3.fr}. The RADMC-3D code is available from \href{https://www.ita.uni-heidelberg.de/~dullemond/software/radmc-3d}{https://www.ita.uni-heidelberg.de/$\sim$dullemond/software/radmc-3d}. The python package used for the post-processing of the simulations data and for the analysis of the radiative transfer calculations are available via its GitHub repository: fargo2radmc3d (\href{https://github.com/charango/fargo2radmc3d}{https://github.com/charango/fargo2radmc3d}). The data obtained in our simulations can be made available on reasonable request to the corresponding author.

\appendix

\section{Higher and lower planet mass}\label{planet-mass}
As described in Section~\ref{sec:hd}, we use the results of \cite{dong2018eq10} -- their Equation 10 -- to derive the mass of the planet on a fixed circular orbit that can reproduce the radial half-width of the gap around 86~au in the HD 163296 disc (this yields $M_{\rm p} \approx 0.26 \mj$ for our disc's aspect ratio). However, as a part of the parameter space study, we also carried out several simulations with $M_{\rm p}$ twice higher or lower than that. The simulations are shown in Table \ref{table1} as NM05H (higher planet mass) for $M_{\rm p} = 0.52 \mj$, and NM05L (lower planet mass) for $M_{\rm p} = 0.13 \mj$. Like Model NM05, these additional models do not allow for planet migration, but we also ran the same two models with planet migration allowed (these are labelled as Models PM05H and PM05L).

\begin{figure*}
\centering
\includegraphics[width=0.49\hsize]{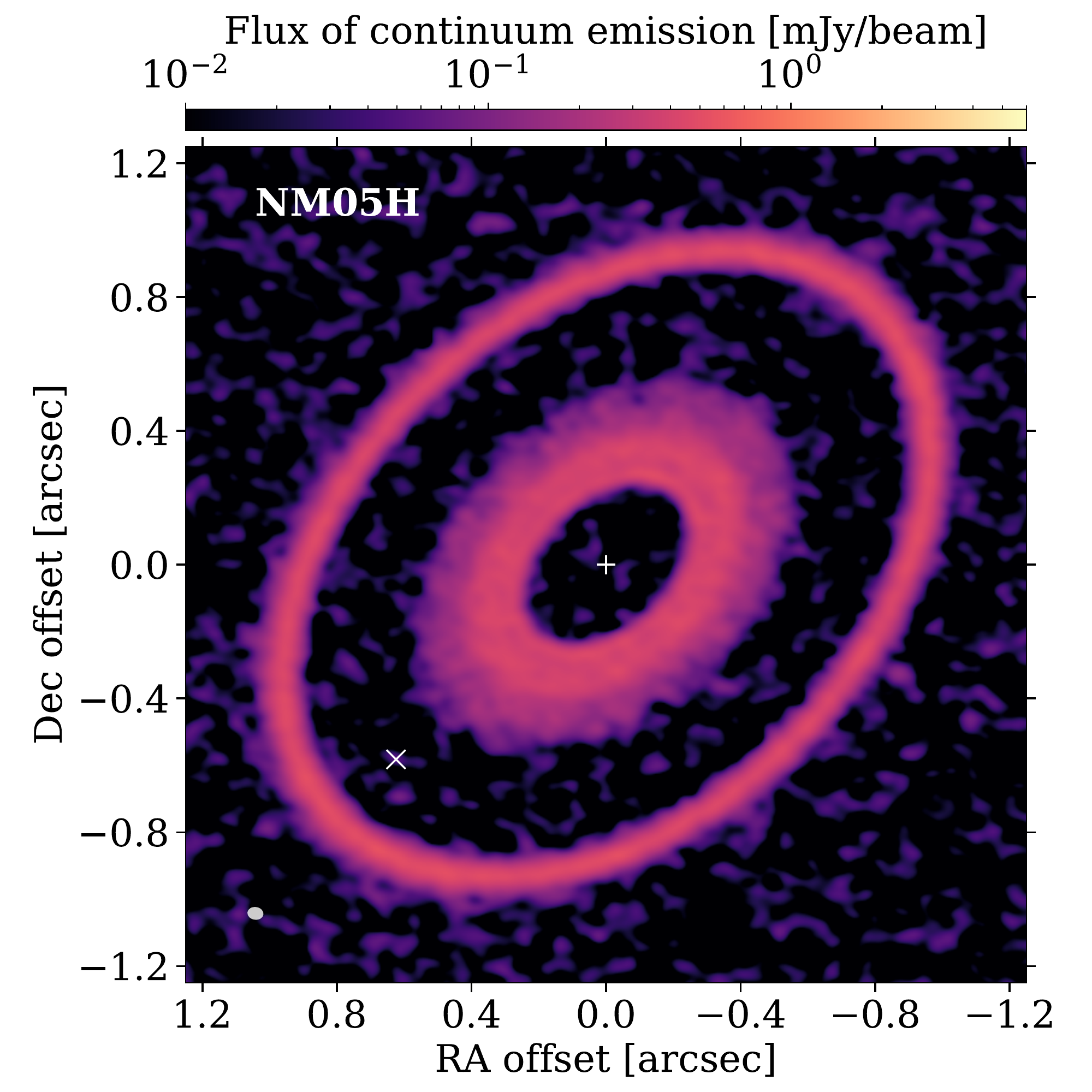}
\includegraphics[width=0.49\hsize]{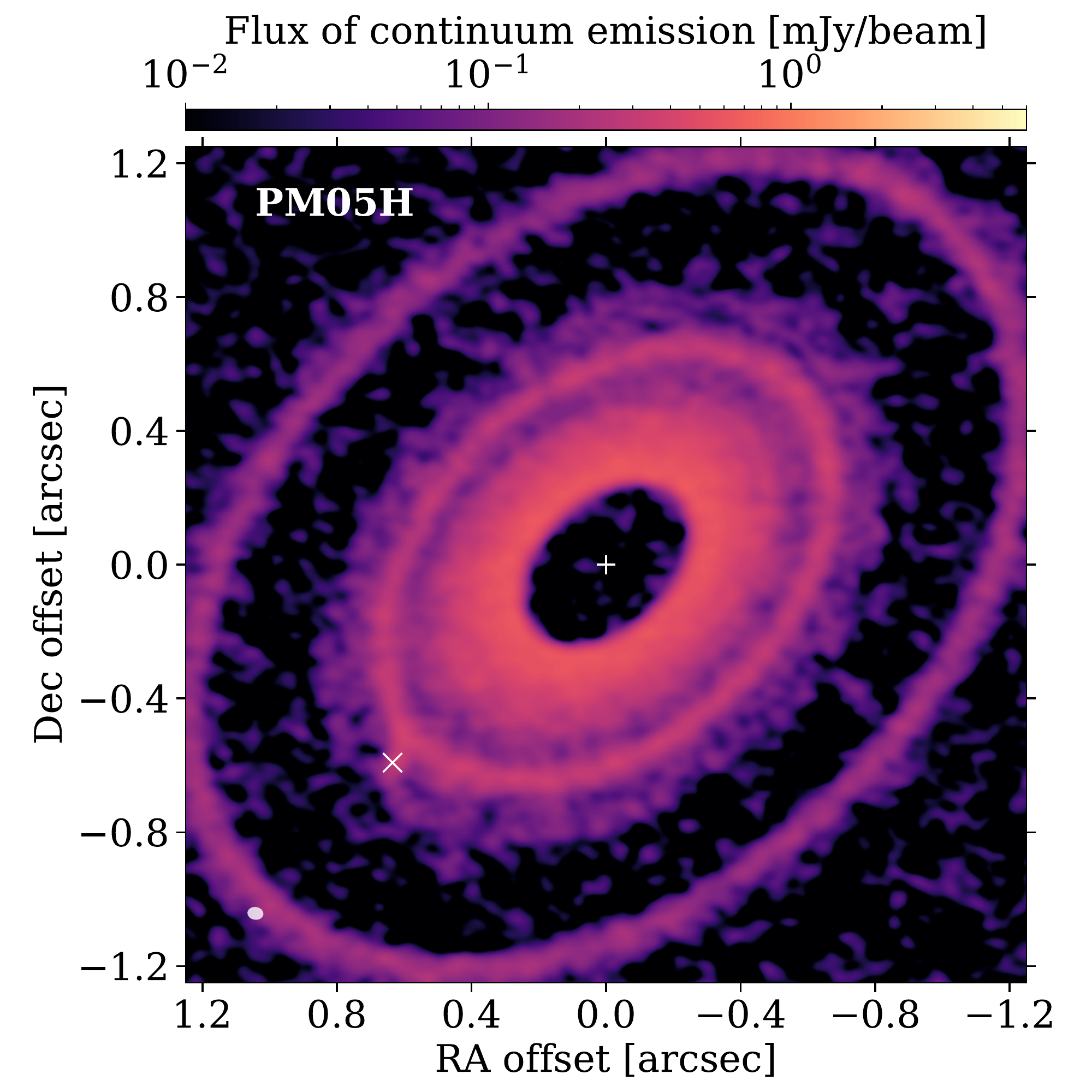}
\includegraphics[width=0.49\hsize]{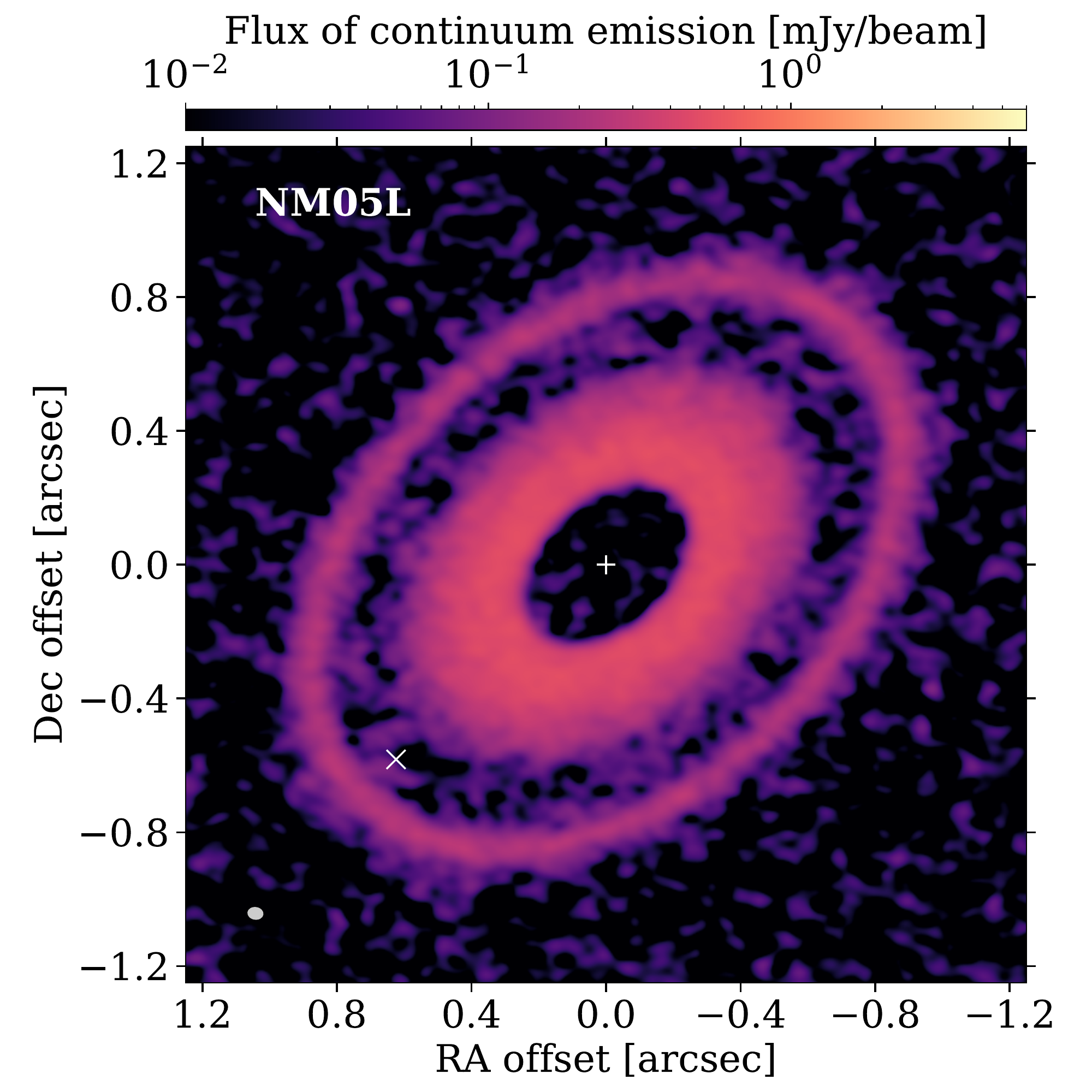}
\includegraphics[width=0.49\hsize]{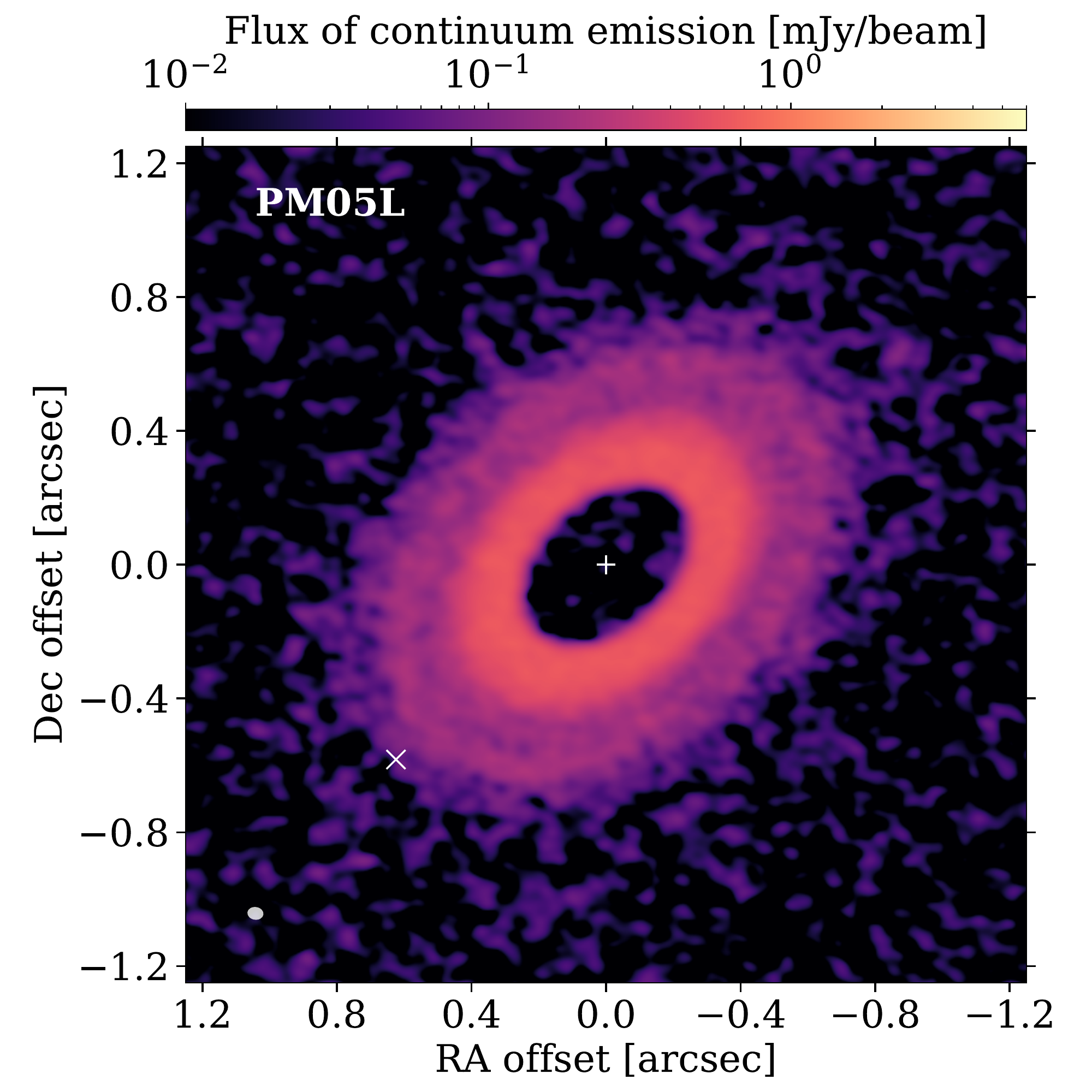}
\caption{\label{fig:different_Mp}Synthetic maps of continuum emission for additional models with different planet masses. In the left column of panels (Models NM05H and NM05L), the planet is held on a fixed circular orbit at 86~au, and the synthetic images are shown at the same time as those in Fig.~\ref{fig:fixed86au}). In the right column of panels, the planet migrates from 130~au and images are shown when it reaches about 86~au (which is at $t=260$~orbits for Model PM05H and $t=620$~orbits for Model PM05L).}
\end{figure*}

Fig. \ref{fig:different_Mp} shows synthetic images of the dust continuum emission for these additional models. For model NM05H (fixed planet, higher mass), the dark ring of continuum emission is significantly wider compared to our fiducial Model NM05 (compare with the top-right panel in Fig.~\ref{fig:fixed86au}). For model NM05L (fixed planet, lower mass), the width of the dark ring is smaller than for Model NM05 and actually in pretty good agreement with the observed one. The width and flux level around the outer bright ring agree somewhat less with the observation, although this could be improved by slightly increasing the dust-to-gas mass ratio. This actually illustrates the degeneracy in the modelling of observed disc structures. Regarding the additional models with migration, the one with higher planet mass (PM05H) undergoes rather fast migration, which causes a very wide dust gap (it takes only 260 orbits for the planet to move from 130~au to 86~au). And finally the model with lower mass (PM05L) shows no dust trapping at the gap's outer edge and therefore no outer bright ring (the larger dust drifts faster than the planet and mostly end up in the inner disc). This further shows how planet migration and dust drift, when both included in the simulations, are able to reduce degeneracy of the models.

\section{A multiple planet scenario in HD 163296}\label{sec:multi-planet}
Although the HD 163296 disc likely features multiple planets, our model, which is inspired from this disc, only has one planet for simplicity. To test how our conclusions would be changed by the presence of several planets, we carried out an additional simulation with multiple migrating planets. More specifically, we used the exact same disc and planet parameters as \cite{Juan2023-HD163296}, where four near-resonant planets were simulated on fixed orbits. These four planets are located at 46~au, 54~au, 84.5~au and 137~au with mass 0.255 $M_{\rm J}$, 0.18 $M_{\rm J}$, 0.4 $M_{\rm J}$ and 1 $M_{\rm J}$, respectively. We modified our source files to adopt the same radial profile of $\alpha$ turbulent viscosity as in Eq.~2 of \cite{Juan2023-HD163296}. The disc mass is 0.15 $\msun$.

Just like \cite{Juan2023-HD163296}, we first ran this simulation for 2000 orbits at 48 au without letting the planets migrate. We have checked that the gas density that we obtain agrees very well with their Figure 1. We note that this requires the gas indirect term to be discarded in the simulation, otherwise the disc gas grows an unstable $m=1$ mode that completely destabilises the disc after a few hundred orbits. We then restarted the simulation by allowing the planets to migrate (planets interact with the disc other and with each other).

\begin{figure}
\begin{centering}
\includegraphics[width=0.49\textwidth]{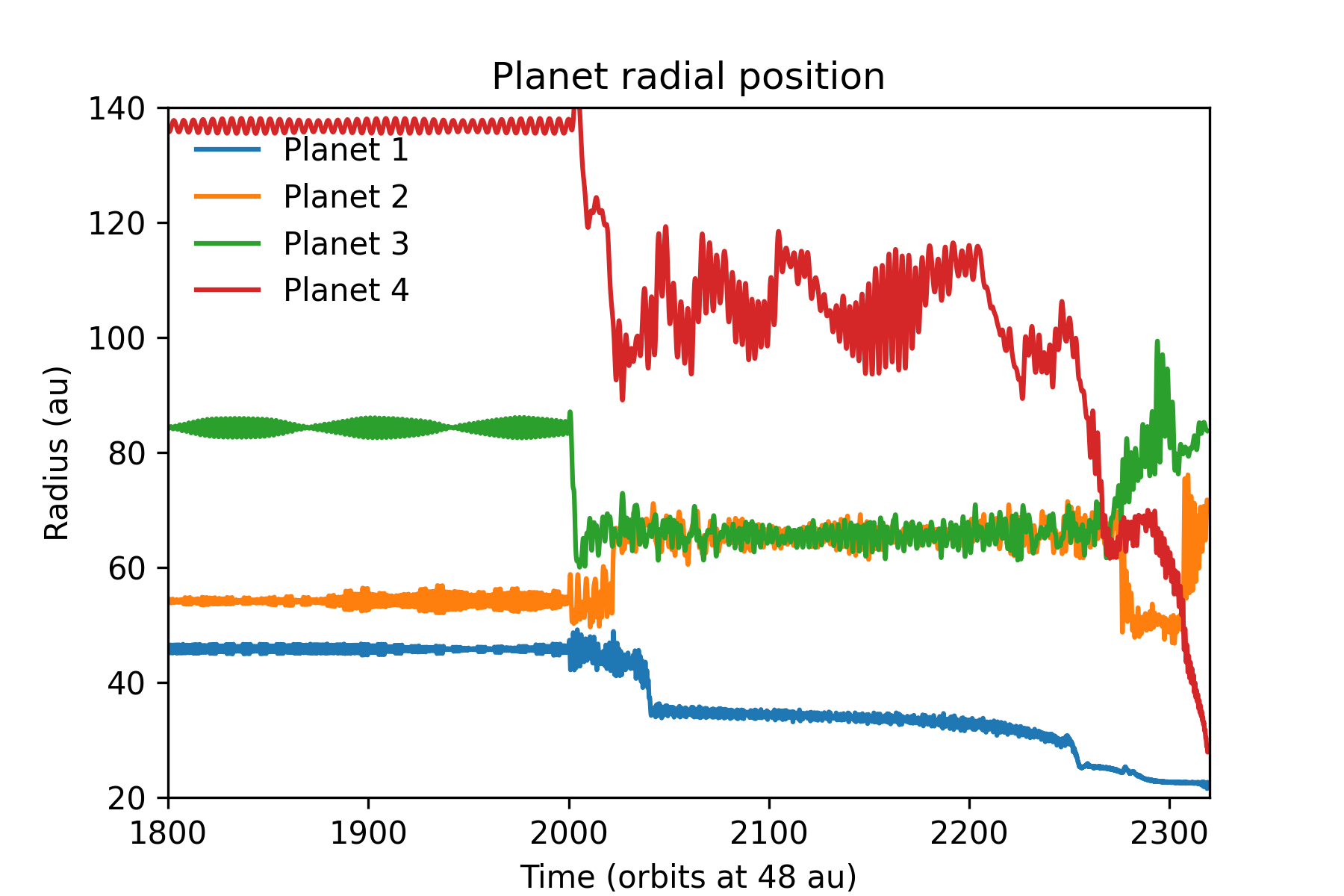}
\par\end{centering}
\caption{\label{fig:multi-planet}Similar to Fig. \ref{fig:migration_rate} but for the four-planet model presented in Sect.~\ref{sec:multi-planet} of the Appendix.}
\end{figure}

The time evolution of the planets orbital radius is displayed in Fig.~\ref{fig:multi-planet}. It is clear that the planets do not maintain their initial (near-resonant) orbits, perhaps not too surprisingly given the high disc mass (much higher than the one from which we get runaway migration with a single planet). Although planet-planet interactions are strong, disc-induced runaway migration remains inevitable, more particularly for the innermost and outermost planets. We point out that the second and third planets temporarily share the same orbital radius for a few hundred orbits while being $\sim 180$ degrees far apart. To our knowledge only a few hydrodynamical simulations of disc-planets interactions have reported that two planets could reach a 1:1 mean-motion resonance through convergent migration; in particular, in the work of \citealp{Penzlin19} for eccentric circumbinary planets, and that of \citealp{Crida09SS}, where a 1:1 resonance is obtained between Uranus and Neptune in a simulation of the four giants in the Solar System that involves a massive version of the Minimum Mass Solar Nebula.

If indeed the HD 163296 disc has several planets near mean-motion resonances, it seems unlikely that the planets acquired such near-resonant states with a high disc mass \citep{Huang2023}. All in all, whether a single planet or multiple planets are invoked, our main conclusions regarding the impact of disc mass on disc migration remain unaltered.

\bibliographystyle{mnras}
\bibliography{ybn_references}

\bsp	
\label{lastpage}
\end{CJK*}
\end{document}